%% file: main.tex
\documentclass{jfm}
\usepackage{graphicx}
\usepackage{graphbox}
\usepackage{newtxtext}
\usepackage{newtxmath}
\usepackage{mathtools}

\usepackage{natbib}
\usepackage{hyperref}
\hypersetup{
    colorlinks = true,
    urlcolor   = blue,
    citecolor  = black,
}

\usepackage{multirow}
\usepackage{accents}
\usepackage{comment}
\usepackage{subcaption}
\usepackage{xcolor} 
\usepackage{epstopdf, epsfig}
\usepackage{tikz}

\newcommand{\RomanNumeralCaps}[1]
\linenumbers
\graphicspath{{figures/}}
\usepackage{cleveref}

\input{imports}

\title{Oscillatory shear flows and a new 2D self-sustaining process}

\author{Theo A. Lewy\aff{1}
 \and Rich R. Kerswell \aff{1}
\corresp{\email{rrk26@cam.ac.uk}} }
  
\affiliation{\aff{1} DAMTP, Centre for Mathematical Sciences, Wilberforce Road, Cambridge CB3 0WA, UK}

\begin{document}

\maketitle

\begin{abstract}
A new self-sustaining process (SSP) is identified  in an oscillating spatially-homogeneous shear flow using a nonlinear Kelvin mode model. This SSP consists of energetic spanwise-invariant 1D `sheets' and weaker spanwise-dependent 2D fluctuations. It is instantaneously 2D, with a time-dependent invariant direction in the flow-cross-shear plane, and has a varicose symmetry. The sheets oscillate in time and are linearly unstable generating fluctuations which then nonlinearly self-interact to reinforce the sheets. The linear instability of the sheet is shown to be due to a lift-up mechanism acting in tandem with `push-forward', which arises due to the `streamwise' shear of the sheets. As the Reynolds number $Re\rightarrow\infty$, both asymptotics and numerics show  SSP lower branches with fluctuation velocity $|\boldsymbol{u}'|\sim Re^{-1/2}$. Implications are discussed for oscillatory wall-bounded flows.

\end{abstract}

%

\section{Introduction}

Periodically-forced fluid flows are ubiquitous in Nature (e.g. tides, cardiovascular flows, respiratory flows) and industry (e.g. gas turbines, internal combustion engines, oscillatory baffled reactors, MEMS micropumps). Being able to predict when the  flow behaviour differs from the simplest possible response is of key concern for flow characteristics such as mixing, heat transfer, wall drag and energy dissipation. They are also of fundamental intellectual interest given they present alternative transition scenarios from more-extensively-studied steadily-forced flows and possibly different turbulent endstates as well given the extra parameter (the Strouhal number) present. 

Historically, the study of fluid flows experiencing purely oscillatory forcing with no mean steady part (the focus here) goes back to Stokes and his so-called second problem where a single, infinite, flat plate bounding an infinite fluid-filled half space oscillates in its plane \citep{Stokes1851}. At low levels of forcing, this causes the  formation of a `Stokes' boundary layer (where the Reynolds number based on the thickness is O(1)) but unravelling what happens at stronger forcing has been a longstanding issue \cite[e.g][]{vonKerczekDavis74, Hall78, BlennerhassettBassom02, BlennerhassettBassom08, Ozdemir14, Biau16, LuoWu10,Gong22, Zhang25, SandovalEaves25}. There are a number of reasons for this which epitomise the difficulty of performing stability analysis for an oscillating flow. Firstly, the periodic time dependence of the base flow requires the application of Floquet theory to discern the fate of small (linear) disturbances which invariably requires a numerical treatment. Secondly, the base flow must undergo periods of stabilizing acceleration and destabilizing deceleration which can give rise to phases of intense energy decay and then growth of perturbations over a period \citep{Biau16}.  This can mean that the flow may become turbulent for only part of the cycle and, separately, give rise to what looks like linear instability but is really a finite amplitude instability triggered by very small but finite disturbances \citep{LuoWu10,Ozdemir14, SandovalEaves25}. As a result of these complications, the stability analysis of periodic flows is understandably far less developed than that for steady flows. In particular, a weakly nonlinear Floquet theory to develop appropriate amplitude equations around a bifurcation point  has yet to be developed. Furthermore, only very recently have efforts been made \citep{SandovalEaves25} to describe mechanisms for how subcritical finite amplitude disturbances could be sustained in such flows. This is in stark contrast  with the situation in steadily-forced shear flows where 3 decades of work is already in place \citep{Hamilton1995, Waleffe1997, Hall10} and garnered much attention \cite[e.g. see the reviews:][]{Kerswell05, Eckhardt07, Kawahara12, Graham21, Avila23}.

Our objective here is to add to the recent findings of \cite{SandovalEaves25} by uncovering another `self-sustaining' mechanism for finite amplitude states in periodically-driven flows where the base response is linearly stable. Whereas \cite{SandovalEaves25} find a spatiotemporally evolving version of the 3D self-sustaining process (SSP) for steadily forced shear flows \citep{Hamilton1995, Waleffe1997, Hall10}, we find a distinctly different time-varying 2D cycle built around 1D sheets interacting with 2D `fluctuation' flow fields. 
A partial explanation for this could be the fact that   we choose to consider perhaps the simplest example of a periodically-forced flow: an unbounded domain  (so no walls as in \cite{SandovalEaves25}) with an imposed oscillating spatially-homogeneous shear motivated by the steady shear model of  Kelvin \citep{Kelvin1887}. In the below, this base flow is initially extended by also enforcing oscillating streaks so that the base flow 
\begin{equation}
    \vb{U_B}= U_B(y,z,t)\vb{e}_x =\left[y + \beta \cos(k_zz)\right]\cos(\omega t)\vb{e}_x,
\label{baseflow}    
\end{equation}
becomes linear unstable at sufficiently large streak amplitude $\beta$ (Kelvin's original model corresponds to $\omega=\beta=0$). The new solution branch is then continued back to $\beta=0$ by homotopy revealing a finite-amplitude, time-dependent state which co-exists with the stable oscillating shear state. 

This approach is not {\em a priori} guaranteed to work but at least the necessary subcriticality of the sinuous streak-induced bifurcation had already been implied by a study of the steady version of the problem  \citep[$\omega=0$, $\beta \neq 0$ in][]{Oxley2025}. These authors studied the transient growth characteristics of streaky flows which have been suggested to be more crucial for sustaining wall-bounded turbulence than their linear instability \citep{Schoppa2002, Lozano21}. As part of that investigation, the authors identified the optimal perturbation (which experiences the most transient growth) and confirmed that it had the correct nonlinear self-interaction to reinforce the imposed streaks (see their §4.8) suggesting the possibility of  self-sustaining cycle. However, pursuing this further to find finite-amplitude solutions in the absence of the imposed streaks is complicated by the fact that the wavenumber in this model ultimately diverges linearly with time under constant shear (see their expression (2.7) or (\ref{ky}) below with $\omega \rightarrow 0$). The oscillatory base flow (\ref{baseflow}) studied here circumvents this issue (again see (\ref{ky})\,) with the hope that the frequency of the oscillation can be taken as small and hence irrelevant for the internal dynamics of the finite amplitude state. In fact, such time scale separation is unfortunately not found  but novel flow states emerge nevertheless.

The paper is structured as follows. The governing equations are introduced in §\ref{formulation} along with the Kelvin mode ansatz used to solve them. §\ref{obtaining the SSP} describes how finite-amplitude periodic orbits are found on a stable temporally-oscillating spatially-constant shear. To do this we identify the instabilities of an oscillating flow augmented with spanwise shear (streaks), and then numerically continue the increasingly nonlinear bifurcating states back to zero streak amplitude. The structure of the periodic orbits is discussed in §\ref{structure of the SSP} by decomposing them into spanwise-independent `sheets' and spanwise-dependent fluctuations. The self-sustaining process (SSP) for these orbits  is found to be closely related to an oscillating sheet solution of the inviscid system which can be analytically determined. In §\ref{mechanisms sustaining} we show that the SSP consists of a two-part process: a linear instability of the inviscid sheets generates fluctuations, and these fluctuations through their nonlinear self-interaction reinforce the sheet. We also perform asymptotics in the $Re\rightarrow \infty$ limit and show that the fluctuations are energised due to lift up of the background shear, and `push forward' of the sheet (this is the streamwise analogue of the cross-shear lift up). A discussion follows in §\ref{discussion}.

%

\section{Formulation} \label{formulation}
We begin by consider an oscillating unidirectional base flow in the (streamwise) $x$ direction with a constant shear in (cross-shear) $y$ and enforced periodic streaky shear in (spanwise) $z$ of the form
\begin{equation}\label{streaky base flow}
    \vb{U_B}= U_B(y,z,t)\vb{e}_x =\left[y + \beta \cos(k_zz)\right]\cos(\omega t)\vb{e}_x,
\end{equation}
where $\beta$ is the dimensionless streak strength, $k_z$ the streak wavenumber and $\omega$ the enforced frequency of the oscillations, non-dimensionalised using the $y-$shear rate, $S$, and the initial fundamental $y-$wavenumber, $k_y$. While the objective of this work is to identify a SSP for  $\beta=0$, (i.e. no imposed streaky shear), this target case is first embedded in the more general $\beta \neq 0$ flow to carry out homotopy, a well-known technique for finding nonlinear solutions \citep[e.g.][]{Nagata90}. The velocity perturbation about the base state (\ref{streaky base flow}), $\u=u\vb{e}_x + v \vb{ e}_y + w\vb{e}_z$ and pressure $p$ then satisfies the Navier-Stokes equation
\begin{align}\label{Navier Stokes perturbation}
    \frac{\partial \u}{\partial t} + \left[y + \beta \cos(k_zz)\right]\cos(\omega t) \frac{\partial \u} {\partial x} + \left[v - \beta k_z w\sin(k_z z)\right]\cos(\omega t)\vb{e}_x \nonumber\\
    + \u \cdot \nabla \u = -\nabla p + \frac{1}{Re} \nabla ^2 \u,
\end{align}
with assumed incompressibility
\begin{equation}\label{incompressibility}
    \nabla \cdot \u = 0.
\end{equation}
Here the Reynolds number is defined as
\begin{equation}
Re\coloneq S/\nu k_y^2
\label{Re}
\end{equation}
where $\nu$ is the kinematic viscosity. Taking $\vb{e}_y \cdot \nabla \cross$ and $\vb{e}_y \cdot \nabla \cross \nabla \cross$ of \cref{Navier Stokes perturbation} and using incompressibility (\ref{incompressibility}) then gives that the cross-shear vorticity $\eta \coloneq \partial u/\partial z-\partial w / \partial x$ and cross-shear velocity $v$ satisfy 
\begin{align}\label{eta equation enforced streak}
\left[ 
\frac{\partial}{\partial t} + 
\left( y  +   \beta \cos(k_z z) \right)
\cos(\omega t)  \frac{\partial} {\partial x} - \frac{1}{Re}\nabla^2
\right]
\eta+ \frac{\partial v}{\partial z} \cos(\omega t) + \beta k_z \sin(k_z z)\cos(\omega t) \frac{\partial v}{\partial y} \nonumber\\  -\beta k_z^2 \cos(k_z z)\cos(\omega t) w +  \frac{\partial}{\partial z}\left( \u \cdot \nabla u \right) 
- \frac{\partial}{\partial x}\left(\u \cdot \nabla w \right) = 0,
\end{align}
\begin{align}\label{v equation enforced streak}
\left[\frac{\partial}{\partial t} + \left[y + \beta \cos(k_zz)\right]\cos(\omega t) \frac{\partial} {\partial x} - \frac{1}{Re}\nabla^2\right] \nabla^2 v + 2\beta k_z \sin(k_z z)\cos(\omega t)\left( \frac{\partial^2 w}{\partial x\partial y} -  \frac{\partial^2 v}{\partial x\partial z} \right) \nonumber\\ 
\hspace{-4cm} -\beta k_z^2 \cos(k_z z)\cos(\omega t) \frac{\partial v}{\partial x} -   \frac{\partial}{\partial y}\left(\nabla \cdot(\u \cdot \nabla \u)\right) + \nabla^2(\u \cdot \nabla v) = 0.
\end{align}

To solve these equations we consider modes with time-dependent wavenumbers - so-called `Kelvin' modes \citep{Kelvin1887} - that naturally deal with the $y$-dependence in the advection term.  Kelvin noticed that the ansatz 
\begin{equation}
[\eta, u, v, w](x,y,z,t) = [\hat \eta, \hat u, \hat v, \hat w](t) \exp \left[i(k_x x + (1-k_x t)y + k_z z)\right]
\end{equation} 
could be used to treat the linear stability problem for a simple unbounded shear (i.e. $U_B=y$ when $\beta=\omega=0$ and where $k_y(0)$ has been used to non-dimensionalise the problem).
The \rm{minimal} extension of this to  accommodate the enforced streaky base flow \citep{Oxley2025} and the nonlinearity of the equations is a double sum of such Kelvin modes of the form
\begin{align}\label{kelvin mode ansatz}
    \eta(x,y,z,t) = \sum_{m=-M}^M \sum_{n=-N}^N \hat \eta_{m,n}(t) \exp [im\left(k_x x + k_y(t)y\right) + ink_z z],
\end{align}
(and similarly for $u,v$ and $w$) where 
\begin{equation}
k_y(t) \coloneq 1-k_x\sin(\omega t)/\omega
\label{ky}
\end{equation}
and $M,N$ control the number of modes included in the truncation. This $k_y(t)$ handles advection by the oscillating shear, while the summation over spanwise modes (the $n$) is required due to the enforced streakiness coupling together the spanwise modes. Lastly, the sum over streamwise and cross-shear harmonics (the $m$) takes account of the nonlinearity. Formally, we could also include a `modulation' parameter in (\ref{kelvin mode ansatz}) as per standard Floquet analysis - e.g. writing $i (n+\tfrac{1}{2}) k_z$ instead of $ink_z$ in the expansion would treat subharmonic spanwise modes as the wavelength of the disturbance would then be twice that of the underlying streaks - and this would be an interesting extension but is not considered here. Instead we concentrate on a harmonic disturbance and the representation (\ref{kelvin mode ansatz}) is minimal in the sense that it is only allows for instantaneously 2D flows, as no variation in the time-varying direction $k_y(t)\vb{e}_x - k_x\vb{e}_y$ is possible. Put another way, the initial state is assumed to have one cross-shear wavenumber and so more a general representation would require a complete set of such wavenumbers. The relevance of this minimal representation will be justified {\em a posteriori} by identifying a  finite-amplitude sustained state within it below.

Proceeding with this minimal ansatz, (\ref{eta equation enforced streak}) and (\ref{v equation enforced streak}) become
\begin{align}\label{eta equation with enforced streak ansatz}
    \frac{d \hat{\eta}_{m,n}}{dt} + \frac{1}{Re}k_{m,n}^2\hat \eta_{m,n}  
    + ink_z \hat v_{m,n}\cos(\omega t)+ \hspace{7cm}\nonumber\\ + \frac{1}{2}\beta\cos(\omega t) \bigg[imk_x(\hat{\eta}_{m,n-1} + \hat{\eta}_{m,n+1}) + mk_yk_z(\hat{v}_{m,n-1} - \hat{v}_{m,n+1}) - k_z^2(\hat{w}_{m,n-1}  
    + \hat{w}_{m,n+1})\bigg]\nonumber\\
    +\sum_{p,q}(\hat u_{m-p,n-q}pk_x + \hat v_{m-p,n-q}pk_y + \hat w_{m-p,n-q}qk_z)(mk_x\hat w_{p,q} - nk_z\hat u_{p,q} ) = 0,
\end{align}
\begin{align}\label{v equation with enforced streak ansatz}
    -k_{m,n}^2 & \frac{d \hat{v}_{m,n}}{dt} + \left[2mk_xk_y \cos(\omega t) - \frac{1}{Re}k_{m,n}^4\right] \hat v_{m,n} \nonumber\\
    &- \frac{1}{2}imk_x\beta \cos(\omega t) \bigg(k_{m,n-1}^2\hat v_{m,n-1} + k_{m,n+1}^2\hat v_{m,n+1}  - 2mk_yk_z(\hat w_{m,n-1} - \hat w_{m,n+1}) \nonumber\\ 
    &\qquad \qquad\qquad\qquad+ 2k_z^2\big[(n-1)\hat v_{m,n-1} - (n+1)\hat v_{m,n+1}\big] + k_z^2(\hat v_{m,n-1} + \hat v_{m,n+1}) \bigg)  \nonumber \\&+ i \sum_{p,q}\left( \hat u_{m-p,n-q}pk_x + \hat v_{m-p,n-q}pk_y + \hat w_{m-p,n-q}qk_z\right) \nonumber \\ &\qquad\qquad \qquad \qquad \times \left[(mk_x\hat u_{p,q} + mk_y\hat v_{p,q} + nk_z\hat w_{p,q})mk_y - k_{m,n}^2 \hat v_{p,q} \right] =0
\end{align}
where $k_{m,n}^2 \coloneq m^2(k_x^2 + k_y^2) + n^2 k_z^2$ and $h_{m,n}^2 \coloneq m^2k_x^2 + n^2 k_z^2$. Using incompressibility and the definition of $\eta$ allows $\hat u_{m,n}$ and $\hat w_{m,n}$ to be found in terms of $\hat v_{m,n}$ and $\hat \eta_{m,n}$ with
\begin{align}\label{kelvin mode u}
    \hat u_{m,n} = \frac{-m^2k_x\ky\hat v_{m,n} - ink_z\hat\eta_{m,n}}{h_{m,n}^2},
\end{align}
and
\begin{align}\label{kelvin mode w}
    \hat w_{m,n} = \frac{-mnk_z\ky\hat v_{m,n} + imk_x\hat\eta_{m,n}}{h_{m,n}^2},
\end{align}
for the cases where $m\neq 0$ or $n\neq 0$. There is a degeneracy associated with the $m=n=0$ mode, where \cref{v equation with enforced streak ansatz} and incompressibility are trivially satisfied, so we set $\hat u_{0,0} = \hat v_{0,0} = \hat w_{0,0}=0$, equivalent to enforcing a zero perturbative mean flow, and hence $\hat\eta_{0,0}=0$ (as $\hat\eta_{m,n} = ink_z\hat u_{m,n} - imk_x\hat w_{m,n} $). 
With this, (\ref{eta equation with enforced streak ansatz}) and (\ref{v equation with enforced streak ansatz}) constitute a nonlinear system of complex ODEs in the $2(2M+1)(2N+1)-2$ complex variables $\{\hat \eta_{m,n}, \hat v_{m,n}\in \mathbb{C}| -M\leq m\leq M , -N\leq n\leq N, (m,n)\neq(0,0)\}$. Enforcing that $\eta$ and $v$ are real with $[\hat \eta_{m,n}, \hat v_{m,n}]=[\hat \eta_{-m,-n}^*, \hat v_{-m,-n}^*]$, where the star denotes a complex conjugate, reduces the number of variables to $D\coloneq(2M+1)(2N+1) - 1$. Equations (\ref{eta equation with enforced streak ansatz}) and (\ref{v equation with enforced streak ansatz}) can then alternatively be written in terms of a state vector ${\phi}\in \mathbb{C}^{D}$ that concatenates these variables, as
\begin{align}\label{ODE form}
    \frac{\partial{{\phi}}}{\partial t} = L(t){{\phi}} + \mathcal{N}(\phi,\phi),
\end{align}
where the linear part $L(t)$ is a matrix with period $2\pi/\omega$, and $\mathcal{N}$ is the nonlinear part. 
We integrate this using an explicit Runge Kutta scheme with stepsizes approximated using 4th order errors and steps accurate to 5th order (RK54 in the python module `scipy').

\subsection{Symmetries}\label{symmetries}
The governing equation (\ref{Navier Stokes perturbation}) is invariant to a spatial shift-reflect symmetry
\begin{equation}
    \Lambda[u,v,w,\eta,p](x,y,z,t) := [-u,-v,w,-\eta,p](-x,-y,z+\pi/k_z,t),
\end{equation}
a spatio-temporal shift-reflect symmetry
\begin{equation}
    \Delta[u,v,w,\eta,p](x,y,z,t) := [-u,v,w,\eta,p](-x,y,z,t+\pi/\omega),
\end{equation}
and a varicose reflection symmetry
\begin{equation}
    \mathcal{R}^e[u,v,w,\eta,p](x,y,z,t) := [u,v,-w,-\eta,p](x,y,-z,t),
\end{equation}
where $u$ and $v$ are even in $z$. While the fully nonlinear equations do \textit{not} satisfy a sinuous reflection symmetry
\begin{equation}
    \mathcal{R}^o[u,v,w,\eta,p](x,y,z,t) := [-u,-v,w,\eta,-p](x,y,-z,t),
\end{equation}
where $u$ and $v$ are odd in $z$, this is recovered when the equations are linearised about $\vb{U_B}$ (see \cref{linear stab of enforced streaky shear}).
A continuous translational symmetry in $x$ exists
\begin{equation}
    T^x_s[u,v,w,\eta,p](x,y,z,t) := [u,v,w,\eta,p](x+s,y,z, t),
\end{equation}
and when the base state has no enforced streaks ($\beta=0$) there is an additional translational symmetry in $z$
\begin{equation}
    T^z_s[u,v,w,\eta,p](x,y,z,t) := [u,v,w,\eta,p](x,y,z+s, t).
\end{equation}
Each of these can be identified as a symmetry by checking the invariance of \cref{Navier Stokes perturbation} under their action, however a final symmetry 
\begin{equation}\label{translational invariance}
    T_s^{x,y}[u,v,w,\eta,p](x,y,z) := [u,v,w,\eta,p](x+k_y(t)s,y-k_xs,z)
\end{equation}
cannot be identified in this way. This can be identified by the invariance of the ansatz (\ref{kelvin mode ansatz}) under its action. Invariance to this symmetry is therefore enforced on the flow by  this minimal representation, meaning it is instantaneously 2D, with variation only allowed in the $k_x\boldsymbol{e}_x+k_y(t)\boldsymbol{e}_y$ and $\boldsymbol{e}_z$ directions - variation in the $k_y(t)\boldsymbol{e}_x-k_x\boldsymbol{e}_y$ direction is not possible.

%
%
\section{SSP Identification}\label{obtaining the SSP}

The existence of a periodic orbit when the base state contains no enforced streaks ($\beta=0$) has to be a manifestation of a  self-sustaining process (SSP) since unbounded spatially-constant shear (oscillating or not) is linearly stable for spatially-periodic disturbances (evident from energy considerations using (\ref{eta equation enforced streak}) and (\ref{v equation enforced streak}) or  Fig.~\ref{linear stability of steaky base} below).  To obtain one such, we first consider the linear instability of the system with imposed streaks $(\beta\neq 0)$, and track a periodic orbit from a bifurcation of the laminar state at finite $\beta=\beta_c>0 $ to $\beta=0$. 

\subsection{Linear stability of the enforced streaky shear $\beta\neq 0$}\label{linear stab of enforced streaky shear}

To consider the linear stability of the enforced streaky shear \cref{streaky base flow}, we linearise (\ref{ODE form}) to obtain 
\begin{equation}\label{linearised ODE}
\frac{\partial{{\phi}}}{\partial t} = L(t){{\phi}}
\end{equation}
where $\phi$ is the state vector consisting of the Kelvin modes and, since $L(t + 2\pi/\omega)=L(t)$, Floquet analysis must be applied. We define $\psi_t(\phi)$ to be the forward time propagation of $\phi$ by $t$ under \cref{linearised ODE}, and the monodromy matrix $\mathcal{M}_{ij}=[\psi_{T}(\vb{e}_j)]_i$, where $[\vb{e}_j]_k=\delta_{jk}$, the Kronecker delta and, $T=2\pi/\omega$. The stability of the system is captured by the eigenvalues $\mu$ of $\mathcal{M}$, corresponding to Floquet multipliers of the system, with instability if $|\mu|>1$.

\begin{figure}
\includegraphics[width=0.9\linewidth]{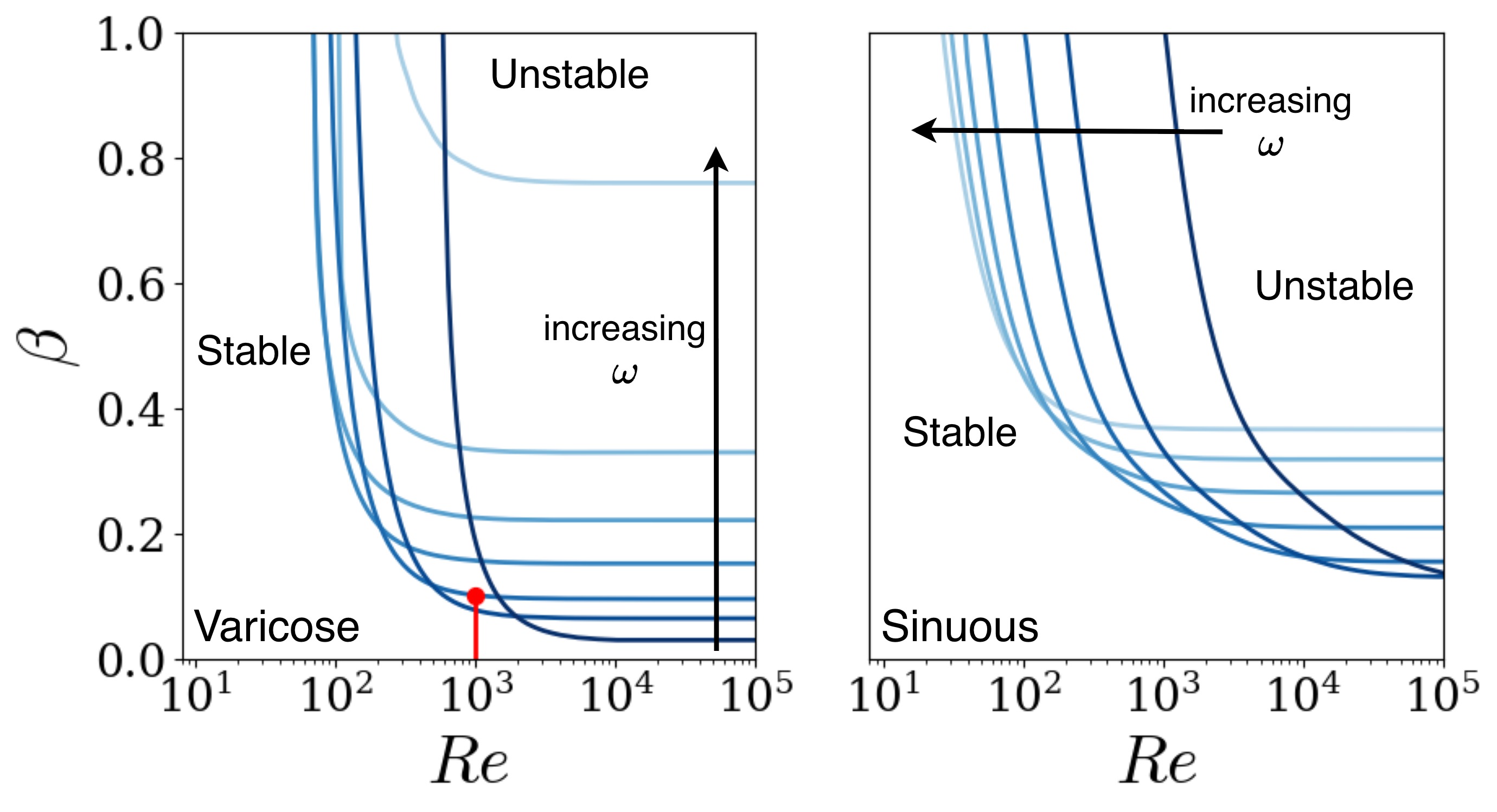} 
\caption{The linear stability across $k_x\in \mathbb{R}$ of the streaky base flow to varicose (left) and sinuous (right) modes, at $k_z=5$, $\omega=0.01,0.05,0.1,0.2,0.3,0.4,0.5$ (dark to light) and resolution $M=N=2$. The red circle and line correspond to the starting point and path taken to continue a periodic orbit down to $\beta=0$ (see \cref{continuation to SSP}). Varicose modes at lower $\omega$ become unstable with only weak enforced streaks (smaller $\beta$). }
\label{linear stability of steaky base}
\end{figure}

The linear problem can be separated into `varicose' modes, which have $\mathcal{R}^e$-symmetry and `sinuous' modes with $\mathcal{R}^o$-symmetry, with their individual neutral curves shown in the $\beta-Re$ plane in  Fig.~\ref{linear stability of steaky base}. Both types of modes show linear instability, with neutral curves pushed to higher $\beta$ as $\omega$ increases, and pushed to higher $Re$ as $\omega$ decreases.  The curves vary only slightly upon increasing the resolution from $M=N=2$ to $M=N=6$, with the varicose critical $\beta_c$ changing by at most 1\%, and the sinuous $\beta_c$ changing by at most 6\% (20\%) when $Re<10^4$ ($Re>10^4$). Varicose modes at low $\omega$  require only weakly enforced streaks (e.g. $\beta=0.1$ at $\omega=0.1$, $Re=1000$) to become linearly unstable. 

%
%

\subsection{Continuation to $\beta=0$}\label{continuation to SSP}

Periodic orbits with period $2\pi/\omega$ were first sought by tracking solutions from the varicose neutral curve of the base state. At $Re=1000$, $\omega=0.1$ and $k_z=5$, the base state first becomes linearly unstable at a critical streak strength $\beta_c = 0.1$ (red dot in Fig.~\ref{linear stability of steaky base}) at the streamwise wavenumber $k_x=0.26$. Starting at this bifurcation of the base state, the branch is subcritical in $\beta$, and we use Newton-GMRES with arclength continuation (see Appendix~\ref{GMRES} for details) to continue the solution down to $\beta=0$. The resultant branch is $\mathcal{R}^e-$symmetric (i.e. varicose).

To find a solution to \cref{ODE form} that is on the desired branch, we first find the unstable Floquet mode just above the neutral curve ${\phi}_c(t)$ 
($=\psi_t(\phi_c(0))$ where $\phi_c(0)$ is the normalised most unstable eigenvector of $\mathcal{M}$). This mode is multiplied by a scale $\alpha$, and then used as an initial guess for a periodic orbit at a $\beta$ just below the neutral curve. We attempted this with $\alpha\in[3\times 10^{-3}, 10^{-2}, 3\times 10^{-2}, 10^{-1}]$, and successfully converged a periodic orbit. This was done to converge 2 solutions on the branch at different $\beta$ just below the neutral curve, and then arclength continuation was used to extend the branch down to $\beta=0$.

\begin{figure}
\centering
\includegraphics[width=0.8\linewidth]{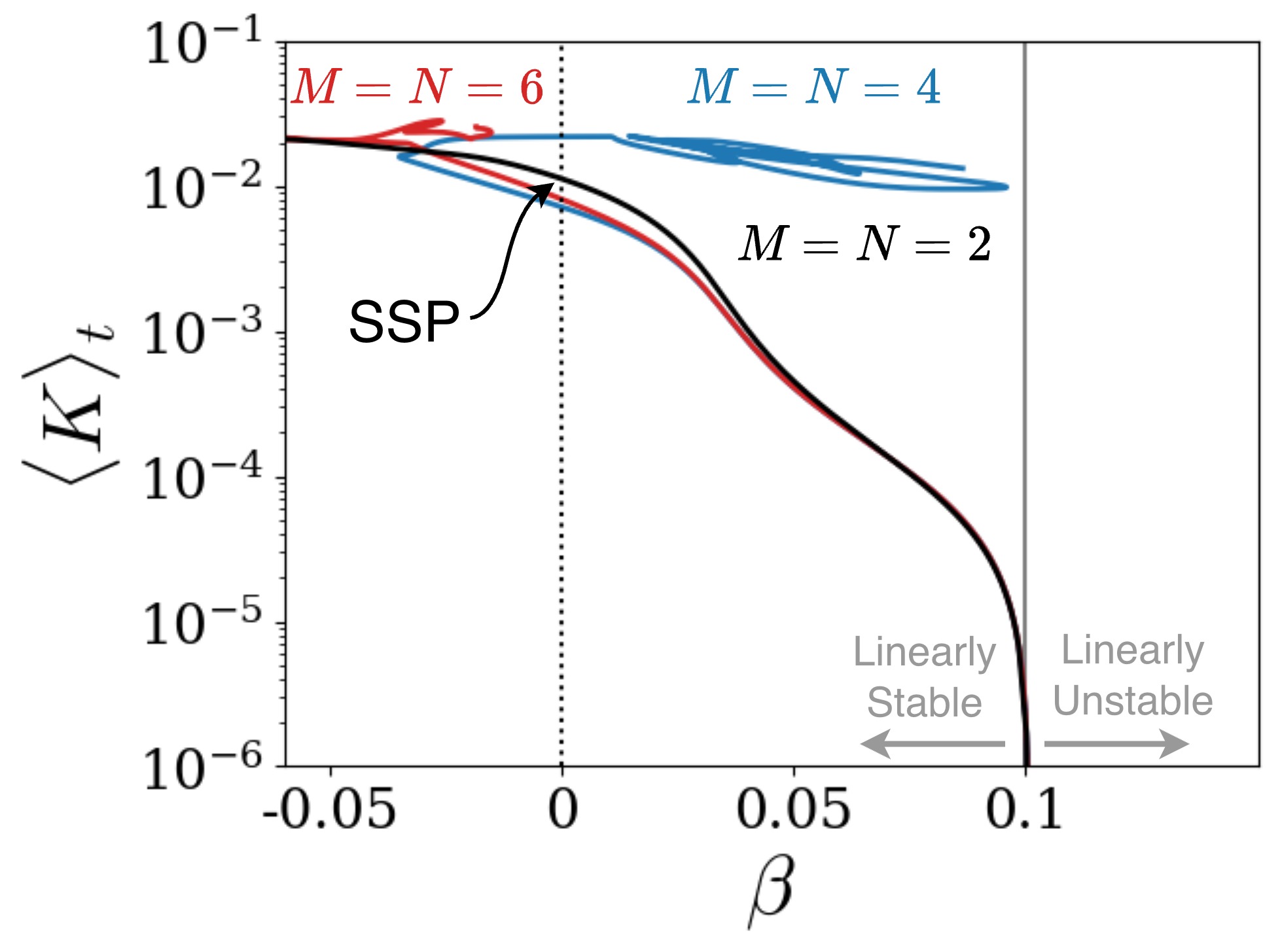} 
\caption{The periodic orbit branch that originates from the linear instability of the base state at $\beta=0.1$ for $Re=1000$, $\omega=0.1$, $k_x=0.26$, $k_z=5$. The time-averaged kinetic energy $\langle K\rangle_t$ is measured at resolutions of $M=N=2$ (black), $4$ (blue) and $6$ (red). A periodic orbit can be found at $\beta=0$, corresponding to a self-sustaining process where the base state is an oscillating simple shear.}
\label{arclength continuation}
\end{figure}

The branch of periodic orbits with period $T=2\pi/\omega$ is shown at different resolutions in Fig.~\ref{arclength continuation}, using the metric of time averaged kinetic energy $\langle K \rangle_t \coloneq\int_0^{T}K dt /T$, where 
\begin{align}\label{KE definition}
    K(\vb{u}) \coloneq \frac{1}{2V_\Omega}\int_\Omega|\u|^2 dV & = \frac{1}{2}\sum_{m,n}(|\hat u_{m,n}|^2 + |\hat v_{m,n}|^2 + |\hat w_{m,n}|^2) \nonumber\\&= \frac{1}{2}\sum_{(m,n)\neq(0,0)}\frac{1}{h_{m,n}}(|\hat \eta_{m,n}|^2 + k_{m,n}^2|\hat v_{m,n}|^2),
\end{align}
with $\Omega\coloneq[-\pi/k_x,\pi/k_x]\times[-\pi,\pi]\times[-\pi/k_z,\pi/k_z]$ and $V_\Omega:= (2\pi)^3/(k_x k_z)$ its volume. The branch extends subcritically all the way back to $\beta=0$ where the base state has no enforced streaks at each resolution checked. This solution therefore corresponds to an SSP in an infinite oscillating shear.  At resolutions of $M=N=2,4,6,8,10$ the SSP has $\langle K\rangle_t=0.0112, 0.0072, 0.0081, 0.0080, 0.0080$, demonstrating convergence with increased resolution (each periodic orbit at $N=M=n$ with $n=4,6,8,10$ was converged by using the orbit at $N=M=n-2$ as an initial guess in the Newton-GMRES solver). In what follows, we use a default resolution of $N=M=2$ unless otherwise specified, which, by retaining only the leading order nonlinearities, corresponds to weakly nonlinear analysis.  

We also attempted to converge a $\beta=0$ periodic orbit from the sinuous modes with $k_z=5$, but  couldn't continue a branch from the neutral curve back to $\beta=0$. For each attempted $Re$, $\omega$, the curve was continued in $\beta$ down from $\beta_c(Re, \omega, k_z)$, which is the smallest (positive) unstable $\beta$, at fixed $k_x = k_{x,c}(Re, \omega, k_z)$ where $k_{x,c}$ is the corresponding marginally unstable $k_x$. The parameter settings $(Re, \omega, \beta_c, k_{x,c})=(10^{3}, 0.1,0.27,0.69), (5\times 10^{4}, 0.05,0.16,0.82),(10^3, 0.2,0.25,1.72)$ were tried with the first being subcritical while the final two were supercritical.

%
%

%
%
\section{The Structure of the SSP}\label{structure of the SSP}

The periodic orbit identified in Fig.~\ref{arclength continuation} consists of spanwise($z$)-invariant 1D `sheets' and spanwise-dependent fluctuations. It is invariant under the $T_s^{x,y}$ symmetry, meaning there is no variation in the time-dependent $k_x\vb{e}_y - k_y(t)\vb{e}_x $ direction. To reflect this, we use a time-dependent rotation of the coordinates
\begin{equation}
    \begin{pmatrix}
        \tilde x(t) \\
        \tilde y(t)
    \end{pmatrix}
    = \Omega(t) \begin{pmatrix}
         x \\
         y
    \end{pmatrix}\coloneq
    \frac{1}{\sqrt{k_x^2+k_y(t)^2}}\begin{pmatrix}
         k_x & k_y(t) \\
         -k_y(t) & k_x 
    \end{pmatrix}
        \begin{pmatrix}
         x \\
         y
    \end{pmatrix},
\end{equation}
and basis vectors
\begin{equation}
    \begin{pmatrix}
        \vb{e}_{\tilde x} \\
        \vb{e}_{\tilde y} 
    \end{pmatrix}
    = \Omega(t) \begin{pmatrix}
        \vb{e}_{ x} \\
        \vb{e}_{ y}
    \end{pmatrix},
\end{equation}
so that the ansatz (\ref{kelvin mode ansatz}) gives rise to an instantaneously 2D velocity field $\vb{u}=\vb{u}(\tilde x,z, t) $. The flow can be decomposed into a spanwise-averaged velocity $\vb{U}({\tilde x},t)$
representing the sheets
\begin{equation}\label{sheet definition}
    \vb{U}(\tilde x,t) \coloneq \langle \vb u\rangle_z = \frac{k_z}{2\pi}\int_{-\pi/k_z}^{\pi/k_z} \vb{u} \, dz = \sum_{m=-M}^M \boldsymbol{\hat u}_{m,0}(t) \exp (im\sqrt{k_x^2 + k_y(t)^2}\, \tilde x(t)),
\end{equation}
and a spanwise-dependent velocity $\vb{u'}({\tilde x},z,t) \coloneq \tilde u'\vb{e}_{\tilde x} + \tilde v'\vb{e}_{\tilde y} +  w'\vb{e}_{z}$ representing the fluctuations
\begin{align}\label{fluctuation definition}
    \vb{u'}(\tilde x,z,t) &\coloneq \vb{u} - \vb{U}\nonumber\\
    &= \sum_{m=-M}^M \mathop{\sum_{n=-N}}_{n\neq0}^N \boldsymbol{\hat u}_{m,n}(t) \exp [im\sqrt{k_x^2 + k_y(t)^2}\, \tilde x(t) + ink_z z].
\end{align}
The sheets therefore correspond to the spanwise-invariant Kelvin modes, $\boldsymbol{\hat u}_{m,0}$, while the fluctuations consist of the spanwise-dependent modes $\boldsymbol{\hat u}_{m,n}$ with $n\neq 0$. Defining $H\coloneq\langle\eta\rangle_z$ and $\eta'\coloneq\eta-H$, the $\mathcal{R}^e$ symmetry of the SSP means $w$ and $\eta$ are odd in $z$ and therefore $W = \langle w \rangle_z =0 $ and $H = \langle \eta \rangle_z =0 $. Further, by incompressibility $\langle {\bf u} \cdot \vb{e}_{\tilde x} \rangle_z =0$, so the flow on the sheets is unidirectional with $\vb{U}(\tilde x, t)=\tilde V(t) \vb{e}_{\tilde y}(t)$ and irrotational with $H=0$. (The standard and the rotated velocity components are related as follows
\begin{equation}
    \begin{pmatrix}
        \tilde u \\
        \tilde v
    \end{pmatrix}
    = \Omega(t) \begin{pmatrix}
         u \\
         v
    \end{pmatrix}, \quad \begin{pmatrix}
        0\\
        \tilde V
    \end{pmatrix}
    = \Omega(t) \begin{pmatrix}
         U \\
         V
    \end{pmatrix}, \quad \begin{pmatrix}
        \tilde u' \\
        \tilde v'
    \end{pmatrix}
    = \Omega(t) \begin{pmatrix}
         u' \\
         v'
    \end{pmatrix} \,\, \, ).
\end{equation}

\begin{figure}
\centering
\includegraphics[width=0.7\linewidth]{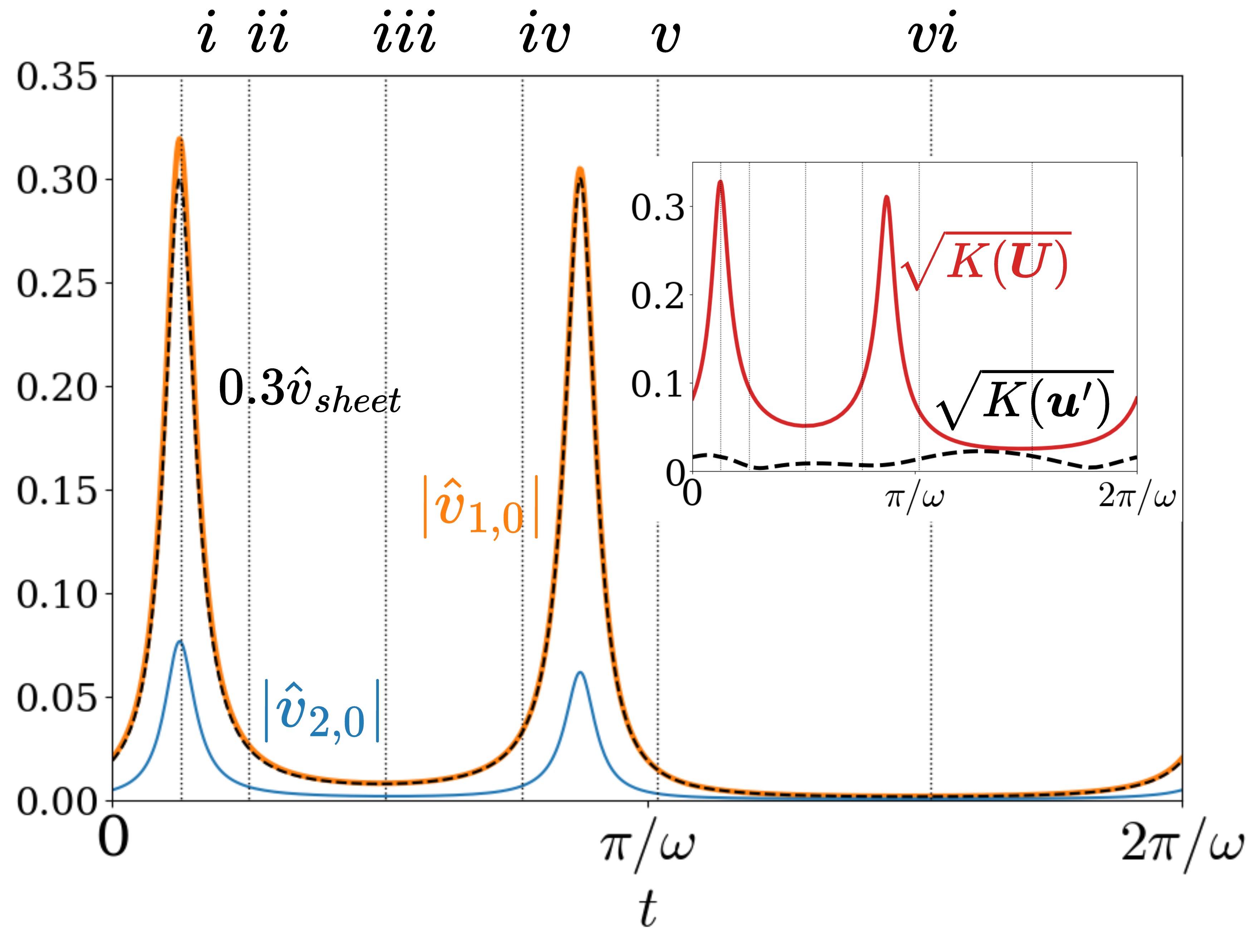} 
\caption{The SSP converged in \cref{continuation to SSP} at $Re=1000$, $\omega=0.1$, $k_z=5$, $k_x=0.26$ with $M=N=2$. The sheet components $|\hat v_{1,0}|$, $|\hat v_{2,0}|$ of the SSP over time alongside an inviscid prediction $0.3\hat v_{sheet}$ (dashed black line, see \cref{approximate sheet}) with kinetic energies $\sqrt{K(\vb{U})}$ and $\sqrt{K({\vb u'})}$ inset. States at times marked $i$-$vi$ are shown in \cref{SSP}b.}
\label{SSP_timeseries}
\end{figure}

\begin{figure}
\centering
\includegraphics[width=0.85\linewidth]{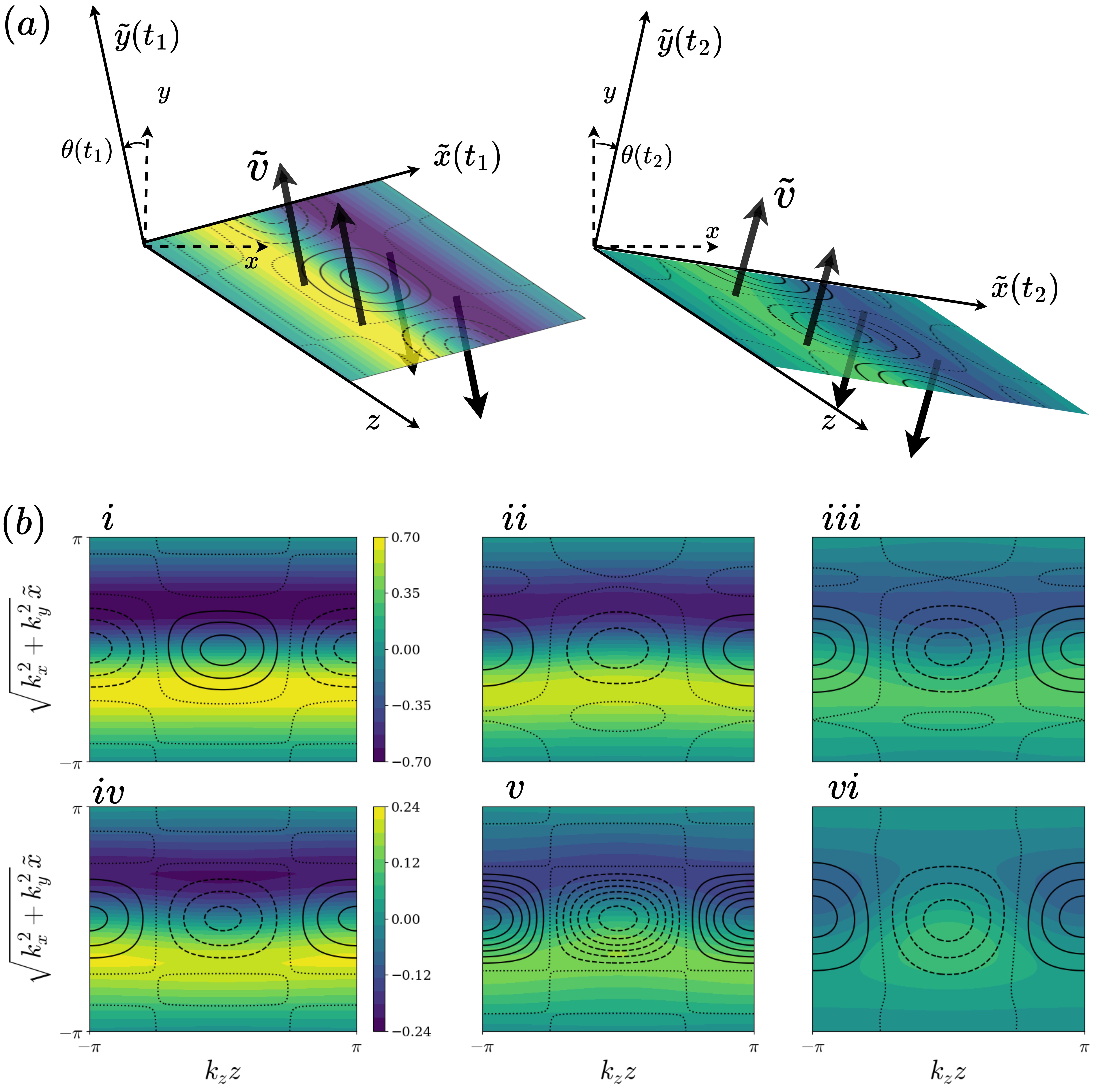} 
\caption{(a) The time-dependent coordinate system shown at different times $t_1,t_2$, with $(\tilde x(t),\tilde y(t))$ the coordinates when $(x,y)$ is rotated about the $z$-axis by an angle of $\theta(t)=\tan^{-1}(k_y(t)/k_x)$. At all times, states are invariant in $\tilde y$, so the whole flow is captured in the $\tilde x$-$z$ planes. Generally, the largest velocity component is $\tilde v$, which is perpendicular to this plane. (b) Velocities in the $\tilde x$-$z$ plane at the times $i$-$vi$ marked in \cref{SSP_timeseries}, with colour contours showing $\tilde v$ (out-of-plane velocity) and line contours showing $\tilde u$ (up-plane velocity), with positive contours solid, negative contours dashed, and zero contours dotted. In $i$ ($ii$-$vi$) each line contour shows a change in $\tilde u$ by $0.02$ ($0.004$). The colour scales for $ii$-$vi$ are the same. Sheets are associated with spanwise independent $\tilde v$, while all other velocities are due to fluctuations.}
\label{SSP}
\end{figure}

Fig.~\ref{SSP_timeseries} shows how the sheet components vary with time for the converged periodic orbit in \cref{continuation to SSP}, with $\sqrt{K(\vb U)}$ and $\sqrt{K(\vb u')}$ plotted in the inset representing the square root of the kinetic energy of the sheet and fluctuation respectively. These show that the sheets are much more energetic than the fluctuations, except around timestamp $(vi)$ where the sheets are `hibernating' and the energy is comparable. Fig.~\ref{SSP}a gives a sketch of the time-dependent $\tilde x$-$z$ plane which captures the whole flow due to invariance in $\tilde y$ and Fig.~\ref{SSP}b shows $\tilde u$ and $\tilde v$ in this plane at various times in the cycle. All states except for $(vi)$ show roughly spanwise invariant $\tilde v$ (see colour contours), as is expected when the sheets are more energetic than the fluctuations.

%
%

%
%

%
%
\subsection{The sheet and fluctuation equations}

The full equations can be decomposed into sheet equations and fluctuation equations by spanwise-averaging. When $\beta=0$, equations (\ref{eta equation enforced streak}) and (\ref{v equation enforced streak}) simplify to
\begin{align}\label{eta with no enforced streak}
    \left[\frac{\partial}{\partial t} + y\cos(\omega t) \frac{\partial} {\partial x} - \frac{1}{Re}\nabla^2\right] \eta+ \frac{\partial v}{\partial z} \cos(\omega t) +  \frac{\partial}{\partial z}\left(\u \cdot \nabla u\right) - \frac{\partial}{\partial x}\left(\u \cdot \nabla w \right) = 0,
\end{align}
\begin{align}\label{v with no enforced streak}
    \left[\frac{\partial}{\partial t} + y \cos(\omega t) \frac{\partial} {\partial x} - \frac{1}{Re}\nabla^2\right] \nabla^2 v 
    -  \frac{\partial}{\partial y}\left(\nabla \cdot(\u \cdot \nabla \u)\right) + \nabla^2(\u \cdot \nabla v) = 0,
\end{align}
and taking a $z$-average then yields the sheet evolution equations
\begin{align}\label{Eta equation}
    \left[\frac{\partial}{\partial t} + y\cos(\omega t) \frac{\partial} {\partial x} - \frac{1}{Re}\nabla^2\right] H - \left\langle \frac{\partial}{\partial x}\left(\u' \cdot \nabla w' \right)\right \rangle_z = 0,
\end{align}
\begin{align}\label{V equation}
    \left[\frac{\partial}{\partial t} + y \cos(\omega t) \frac{\partial} {\partial x} - \frac{1}{Re}\nabla^2\right] \nabla^2 V 
    -   \left\langle \frac{\partial}{\partial y}\left(\nabla \cdot(\u' \cdot \nabla \u')\right) - \nabla^2(\u' \cdot \nabla v') \right\rangle_z = 0,
\end{align}
where terms quadratic in the sheet velocity vanish as $\vb{U}\cdot \nabla \vb{U} = \tilde V \partial_{\tilde y}\vb{U}(\tilde x,t) = 0$. The varicose symmetry of the SSP means equation (\ref{Eta equation}) is trivially satisfied by $H=0$. Subtracting \cref{Eta equation} from \cref{eta with no enforced streak} and subtracting \cref{V equation} from \cref{v with no enforced streak} gives the fluctuation evolution equations
\begin{align}\label{eta fluctuations equations}
    \left[\frac{\partial}{\partial t} + y\cos(\omega t) \frac{\partial} {\partial x} - \frac{1}{Re}\nabla^2\right] \eta'+ \frac{\partial v'}{\partial z} \cos(\omega t) + \frac{\partial}{\partial z}\left(\u' \cdot \nabla U \right) \hspace{2cm}\nonumber \\ 
    +  \frac{\partial}{\partial z}\left(\u' \cdot \nabla u'\right) - \frac{\partial}{\partial x}\left(\u' \cdot \nabla w' \right) + \left\langle \frac{\partial}{\partial x}\left(\u' \cdot \nabla w' \right)\right \rangle_z= 0,
\end{align}
\begin{align}\label{v fluctuations equations}
    \left[\frac{\partial}{\partial t} + y \cos(\omega t) \frac{\partial} {\partial x} - \frac{1}{Re}\nabla^2\right] \nabla^2 v' 
     + \nabla^2(\u' \cdot \nabla V) \hspace{4cm}\nonumber\\
    -   \frac{\partial}{\partial y}\left(\nabla \cdot(\u' \cdot \nabla \u')\right) + \nabla^2(\u' \cdot \nabla v') 
    +\left\langle \frac{\partial}{\partial y}\left(\nabla \cdot(\u' \cdot \nabla \u')\right) - \nabla^2(\u' \cdot \nabla v') \right\rangle_z
    = 0,
\end{align}
where $\vb{U}\cdot \nabla \vb{u}' = \tilde V \partial_{\tilde y}\vb{u}'(\tilde x,z,t) = 0$ and $\nabla \cdot (\vb u ' \cdot \nabla \vb U) = \nabla \cdot (\vb u ' \cdot \nabla \tilde V\vb{e}_{\tilde y}) = \partial_{\tilde y}(\vb u ' \cdot \nabla \tilde V)=0$ were used for simplification. For varicose modes, the sheet equation (\ref{V equation}) and fluctuation equations (\ref{eta fluctuations equations}), (\ref{v fluctuations equations}) define the full system to be solved by $\eta$ and $v$ Kelvin modes (with the $u$ and $w$ velocity components slaved to these by (\ref{kelvin mode u}) and (\ref{kelvin mode w}).

\subsection{Exact inviscid sheet solutions}\label{approximate sheet}

We now identify how sheets evolve in the absence of fluctuations ($\vb u'=\eta'=0$) and find an exact solution in the inviscid limit. Fluctuation equations (\ref{eta fluctuations equations}) and (\ref{v fluctuations equations}) are trivially solved with $\vb u'=\eta'=0$, while \cref{V equation} becomes
\begin{align}\label{exact sheet equation}
    \left[\frac{\partial}{\partial t} + y \cos(\omega t) \frac{\partial} {\partial x} - \frac{1}{Re}\nabla^2\right] \nabla^2 V = 0,
\end{align}
where the sheets are only advected by $\vb U_B=y\cos(\omega t)\vb{\hat x}$, and damped by viscosity. 
Substituting $V=\sum_{m=-M}^M {\hat v}_{m,0}(t) \exp [im(k_x x + k_y y)]$ from \cref{sheet definition} into (\ref{exact sheet equation}) then gives
\begin{align}
    \frac{d}{dt} \biggl( \left[k_x^2+k_y(t)^2 \right] \hat{v}_{m,0}\biggr)
     + \frac{m^2\left[k_x^2 +k_y(t)^2\right]^2}{Re}\hat v_{m,0}=0,
\end{align}
for each $m$. After dividing by $\left[k_x^2 + k_y(t)^2\right]\hat v_{m,0}$ and integrating we obtain
\begin{align}\label{Exact Sheet 1}
\hat v_{m,0} = \frac{A_m}{k_x^2 + k_y^2}\exp\left(-\frac{m^2}{Re}\left[(k_x^2+1)t + \frac{2k_x}{\omega^2}\cos(\omega t) + \frac{k_x^2}{\omega^2}\left(\frac{t}{2} - \frac{\sin(2\omega t)}{4\omega}\right)\right]\right),
\end{align}
with $A_m$ an undetermined amplitude. In the inviscid system ($Re=\infty$) the secular viscous decay is suppressed, and each mode has an exact periodic solution
\begin{align}\label{Exact Sheet inviscid}
\hat v_{m,0} = \frac{A_m}{k_x^2 + k_y(t)^2},
\end{align}
or in terms of the full sheet 
\begin{align}\label{Exact Sheet inviscid V}
{\tilde V}=\sum_{m=1}^M \frac{A_m}{k_x^2 + k_y(t)^2} \exp (im\sqrt{k_x^2 + k_y(t)^2}\, \tilde x)+c.c.
\end{align}

We expect that these inviscid sheet solutions derived in the absence of fluctuations may form the backbone of the periodic orbit  when both small viscosity and fluctuations are present.  To corroborate this, the  evolution of the simple inviscid solution 
\begin{align}\label{Exact Sheet inviscid 2}
\hat v_{sheet} = \frac{k_x^2}{k_x^2 + k_y(t)^2},
\end{align}
is readily fitted to match the observed evolution of the sheet components of the converged SSP: see in Fig.~\ref{SSP_timeseries} that the dashed black line indicating $0.3\hat v_{sheet}$ is almost identical to the shape of $|\hat v_{1,0}|$. 

%
%
\section{Mechanism sustaining the SSP}\label{mechanisms sustaining}

In the absence of fluctuations, we have seen that the sheets oscillate with frequency $T=2\pi/\omega$ due to advection by the base flow $\vb U_B$, while decaying secularly due to viscosity (see (\ref{Exact Sheet 1})\,). To form a periodic orbit, the SSP must overcome the secular decay and it does this via two processes. Firstly, the oscillating sheets are linearly unstable, generating fluctuations and, secondly, these fluctuations interact nonlinearly, re-energising the original sheets. To verify this, we converge multiple branches of SSPs by considering different modes of instability of the sheets.

\subsection{Linear instability of the sheets produces fluctuations}\label{linear instab of sheet generate fluc}

The linear stability of the oscillating sheet solution is determined by linearising (\ref{eta fluctuations equations}) and (\ref{v fluctuations equations}) which gives
\begin{align}\label{eta fluctuations equations linearised}
    \left[\frac{\partial}{\partial t} + y\cos(\omega t) \frac{\partial} {\partial x} - \frac{1}{Re}\nabla^2\right] \eta'+ \frac{\partial v'}{\partial z} \cos(\omega t) + \frac{\partial}{\partial z}\left(\u' \cdot \nabla U \right)
    = 0,
\end{align}
\begin{align}\label{v fluctuations equations linearised}
    \left[\frac{\partial}{\partial t} + y \cos(\omega t) \frac{\partial} {\partial x} - \frac{1}{Re}\nabla^2\right] \nabla^2 v' 
    + \nabla^2(\u' \cdot \nabla V) =0.
\end{align}
Writing the fluctuations in modal form (\ref{fluctuation definition}) and enforcing real $\eta,v$ yields a linear system of complex ODEs (see equations (\ref{eta equation with enforced sheets}) and (\ref{v equation with enforced sheets}) in Appendix \ref{equations}). As both the base shear and the oscillating sheets have period $2\pi/\omega$, this takes the form of \cref{linearised ODE}, allowing us to use Floquet analysis to consider its stability.

We now choose a family of oscillating sheets to investigate. Using resolution $M=2$ with (\ref{sheet definition}) and (\ref{Exact Sheet inviscid}), we obtain the general oscillating sheet
\begin{align}\label{specific sheets}
    {\tilde V}_{A,B} &= \sum_{m=-2}^2 {\hat v}_{m,0}(t) \exp (im\sqrt{k_x^2 + k_y^2}\, \tilde x) \nonumber\\
    &=A\left(\frac{1}{k_x^2+k_y^2}\exp(i\sqrt{k_x^2 + k_y^2}\, \tilde x) + \frac{B}{k_x^2+k_y^2}\exp(2i\sqrt{k_x^2 + k_y^2}\, \tilde x)\right) + c.c. 
\end{align}
where ${\hat v}_{0,0}(t)=0$ by the discussion following (\ref{kelvin  mode w}), the constant $A$ controls the amplitude of the $m=1$ mode, the constant $B$ controls the ratio of the amplitudes of the $m=1$ and $m=2$ modes, and the $c.c.$ denotes its complex conjugate. We set $B=-0.123$, motivated by time-averaging the ratio $\hat v_{2,0}/\hat v_{1,0}$ of the SSP in \cref{SSP_timeseries}, leaving us with a single `sheet-strength' parameter $A$. The streamwise sheet velocity $U_{A,B}$ can be recovered using \cref{kelvin mode u} and the spanwise velocity $W_{A,B}=0$ by symmetry as commented on earlier, giving the full sheet velocity $\vb U_{A,B}$.

We perform Floquet analysis about these sheets by solving equations (\ref{eta fluctuations equations linearised}) and (\ref{v fluctuations equations linearised}) with $\vb U=\vb U_{A,B}$ as the sheet-strength $A$ is varied, and show the resultant growth rates in \cref{linear stability of sheets}. Multiple unstable modes exist.

%
%
\begin{figure}
\centering
\includegraphics[width=0.85\linewidth]{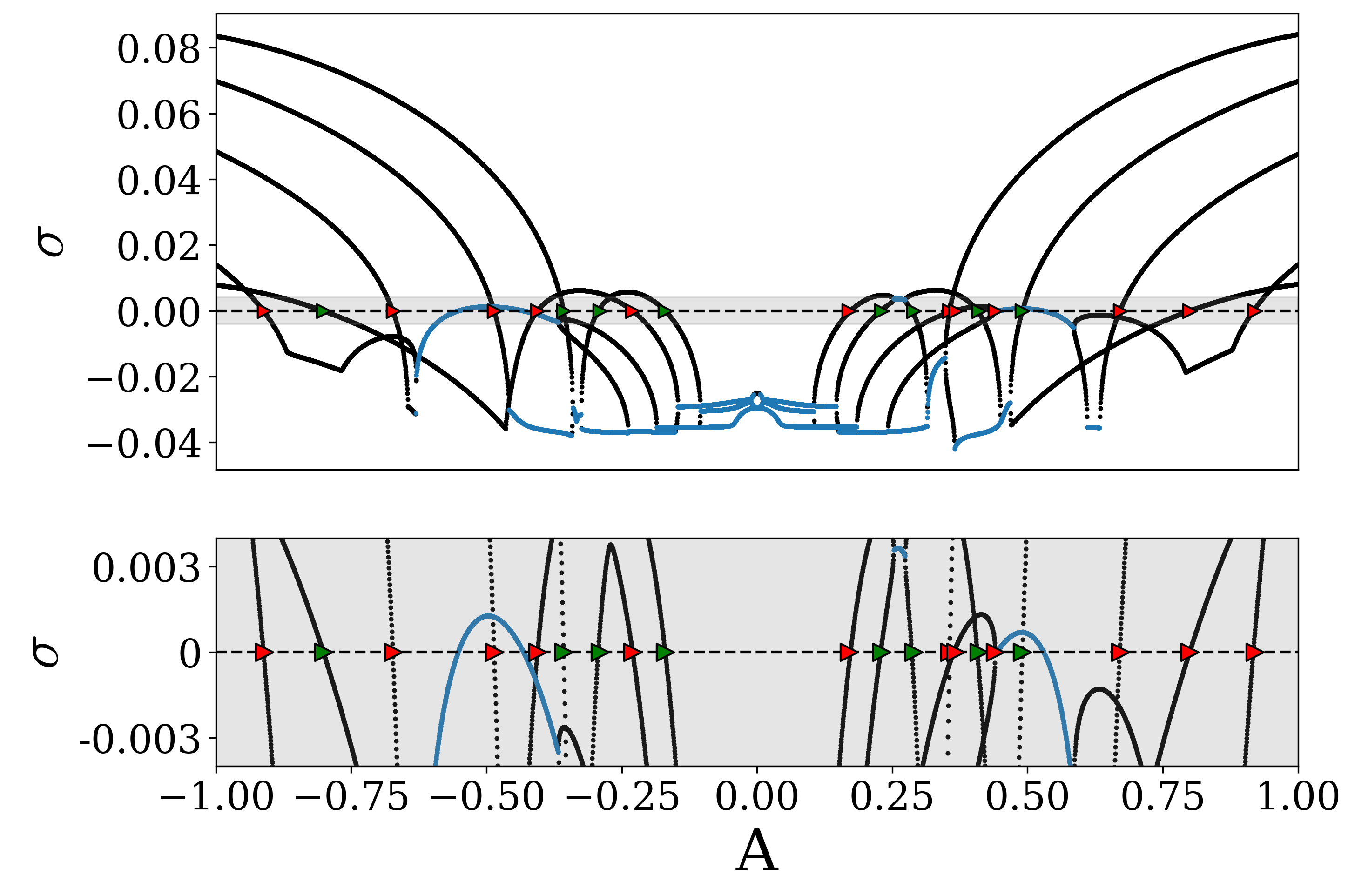} 
\caption{Linear stability of sheets \cref{specific sheets} as sheet-strength $A$ varies, showing the 5 most unstable growth rates $\sigma=\log(|\mu|) /T$ where $\mu$ is the Floquet multiplier. Parameters are $Re=1000$, $\omega=0.1$, $k_z=5$, $k_x=0.26$ with $M=N=2$. Black (blue) shows real (complex) $\mu$. Green (red) symbols show where neutral eigenmodes succeeded (failed) in allowing the convergence of an SSP (see \cref{reinforce}). The bottom figure shows a zoom-in of the grey stripe in the top figure. This demonstrates that sheets are unstable to many modes of instability.}
\label{linear stability of sheets}
\end{figure}

%
%
\begin{figure}
\centering
\includegraphics[width=0.85\linewidth]{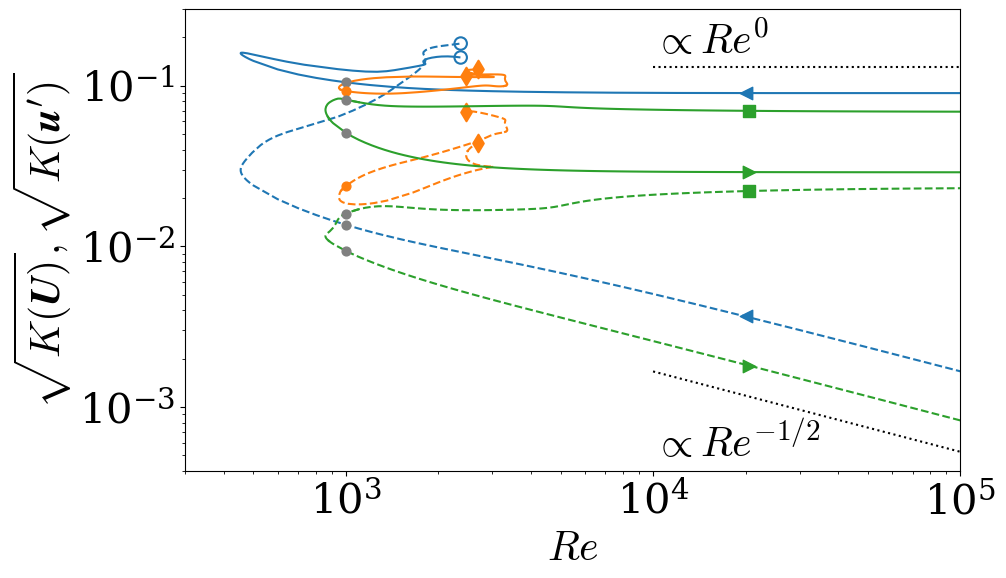} 
\caption{Continuation in $Re$ of the converged orbits found using the protocol described in \cref{reinforce} (the starting SSPs are marked with filled circles at $Re=1000$ - note one orbit corresponds to two circles, one on the sheet energy curve and one on the fluctuation energy curve). Parameters are $\omega=0.1$, $k_z=5$, $k_x=0.26$ with $M=N=2$. Branches show $\sqrt{K(\vb{U})}$ (solid) and $\sqrt{K(\vb{u'})}$ (dashed). The blue branch corresponds to the SSP identified in \cref{continuation to SSP} with period $T=2\pi/\omega$, while the others each have $T=4\pi/\omega$. The open blue circle shows where the blue branch turns around and retraces itself. The orange diamonds show where the orange branch was unable to be continued by a step of $\delta Re>1$. Dotted black lines show that blue and green branches have the predicted asymptotic scaling of $|\vb U|\sim Re^0$ and $|\vb u'|\sim Re^{-1/2}$. Sheet components $|\hat v_{m,0}|$ for the filled coloured symbols are shown in \cref{New SSP Sheets}}
\label{Fluctuation and Sheet scales}
\end{figure}

%
%
\begin{figure}
\centering
\includegraphics[width=0.85\linewidth]{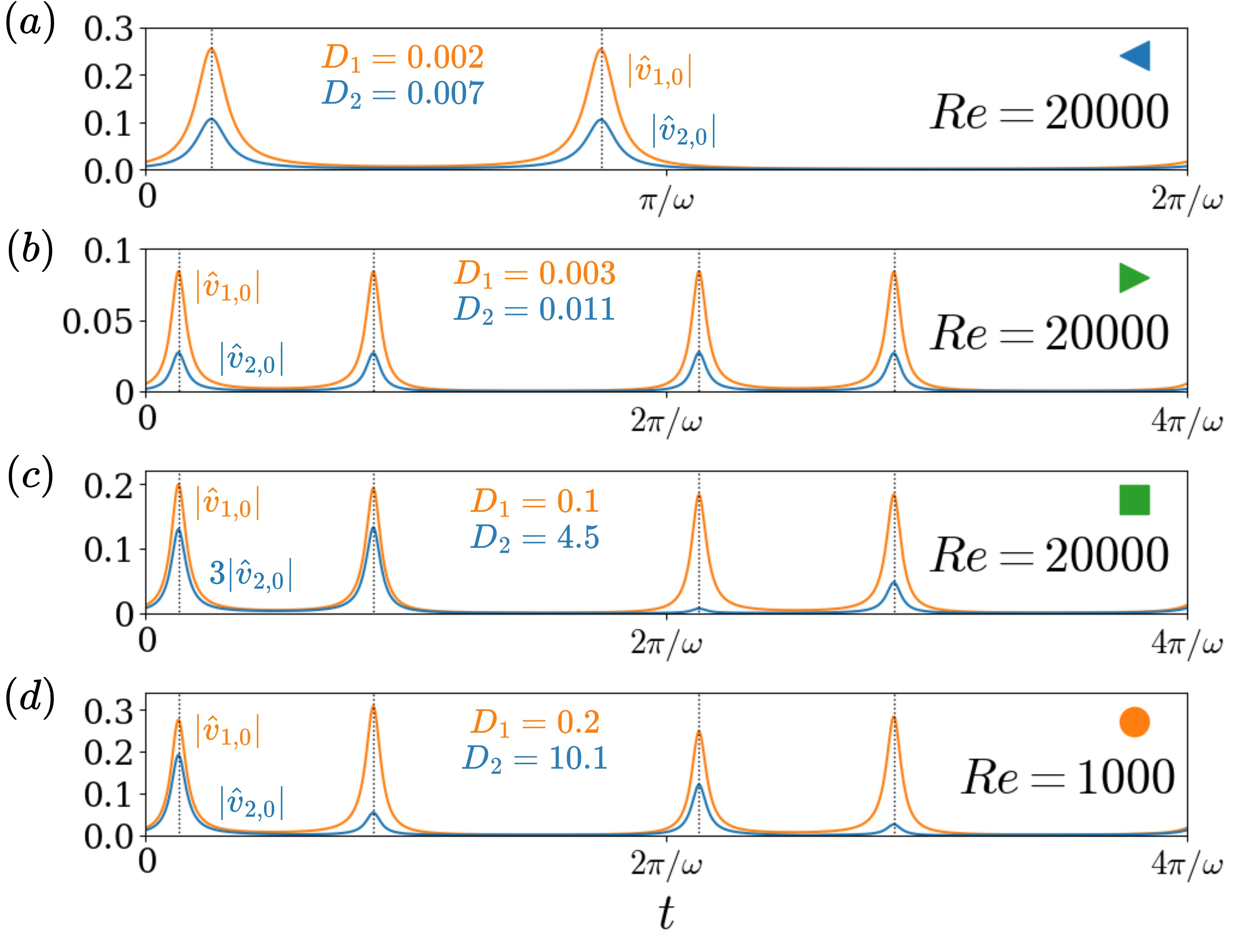} 
\caption{The sheet components $|\hat v_{1,0}|$ and $|\hat v_{2,0}|$ for a range of symbols shown in \cref{Fluctuation and Sheet scales}. Dotted vertical lines show where $k_y(t)=0$, which is where inviscid sheets reach their maxima. $D_i$ quantifies the deviation of each sheet component from an inviscid solution. As $Re\rightarrow \infty$, (a) and (b) correspond to SSP where fluctuations scale like $|\vb{u}'|\sim Re^{-1/2}$ and the sheets are asymptotically inviscid (see \cref{asymptotics} for asymptotics). (c) and (d) show SSP that do not have these asymptotics. }
\label{New SSP Sheets}
\end{figure}

%
%
\subsection{Fluctuations nonlinearly reinforce sheets}\label{reinforce}

For equation (\ref{V equation}) to be solved, the nonlinear interactions of the fluctuations $\eta',v'$ must be of the right form to reinforce the sheet $V_{A,B}$, countering the viscous dissipation. One way to see if these fluctuations are likely to reinforce the sheets is to see what shape of sheets are induced in (\ref{V equation}) when the nonlinear fluctuations force the system, as was originally considered for the regeneration of rolls in the roll-streak-wave  SSP of a steady shear \citep{Waleffe1997}. A more definitive approach, however, is instead to construct guesses using the  sheet and neutral eigenfunctions (fluctuations)  for a periodic orbit of the full system of equations (\ref{V equation}), (\ref{eta fluctuations equations}) and (\ref{v fluctuations equations}), and then try to converge it in the $\beta=0$ equations using Newton-GMRES (see Appendix \ref{GMRES}).

The guesses were constructed as $v=V_{A,B}+\alpha v'$ and $\eta=\alpha \eta'$ with $B=-0.123$, using bifurcation values of $A$ where a Floquet mode is neutral ($\sigma=0$ in \cref{linear stability of sheets}) with $\eta',v'$ the corresponding normalised Floquet mode. The fluctuation amplitude $\alpha$ is unknown so a range of values were tried from $\alpha=0.01$ to $0.09$ in steps of $0.01$. Only neutral Floquet modes with multiplier $\mu=\pm 1$ were used (i.e. the blue neutral modes in \cref{linear stability of sheets} were ignored), as these produce simpler harmonic or subharmonic periodic orbits that repeat exactly after one forcing period ($T=2\pi/\omega$) or two ($T=4\pi/\omega$) respectively rather than complex multipliers that produce recurrent periodic orbits (which only repeat after $T$ time followed by a spatial translation). Importantly, Floquet analysis does not allow orbits of smaller period than the underlying sheets or base shear to be generated. So this constructive approach can never produce orbits with periods much smaller than that of the base shear whereas the opposite is possible, that is, periodic orbits of much longer period than the base shear.

The green(red) symbols in \cref{linear stability of sheets} show where this protocol succeeds(fails) in converging an SSP at the different $A$ tried.
Of the 8 successes, 3 distinct solution curves were found due to multiple convergences to the same curve and degeneracies due to symmetries. The
sheet and fluctuation kinetics energies of these periodic orbit curves as a function of $Re$ are plotted in Fig.~\ref{Fluctuation and Sheet scales} with the blue line indicating a harmonic ($T=2\pi/\omega$) periodic orbit and the orange and green curves indicating subharmonic periodic orbits ($T=4\pi/\omega$). All periodic orbit solution curves have saddle node bifurcations at a minimum $Re$ where upper and lower branches meet. For the green curve which can be fully continued to high $Re$, the kinetic energy of the fluctuations stays $O(1)$ as $Re \rightarrow \infty$ on the upper branch whereas it decays like $O(Re^{-1})$ on the lower branch.

We show the sheet oscillations of these periodic orbit branches in \cref{New SSP Sheets}. In all cases the SSP shows spikes in $|\hat v_{m,0}|$ for $m=1,2$ when $k_y(t)=0$ like the inviscid solution (\ref{Exact Sheet inviscid}). This temporal behaviour of $|\hat v_{m,0}|$ can be compared to that of the  inviscid solution $A\hat v_{sheet}$ which has the same maximum ($A = \max_t(|\hat v_{m,0}|)$). Measuring the deviation from this inviscid solution using $$D_m=\frac{\omega}{2\pi}\int_0^{2\pi/\omega}\frac{\left||\hat v_{m,0}| - A\hat v_{sheet}\right|}{|\hat v_{m,0}|}\, dt, \quad \text{with} \,A = \max_t(|\hat v_{m,0}|) $$
then  $D_m \ll 1$ indicates that  the $\hat v_{m,0}$ sheet mode is close to an exact inviscid oscillating sheet. The sheets of \cref{New SSP Sheets}(a) and (b) have $D_m<0.012$ for $m=1,2$, demonstrating the relevance of the inviscid solution. In contrast, (c) and (d) have $D_1=O(0.1)$ and $D_2>1$, suggesting that the harmonic sheet mode ($m=1$) is roughly inviscid but the same is not true for the subharmonic ($m=2$).

%
%
\subsection{Asymptotics as $Re\rightarrow \infty$}\label{asymptotics}

Of the three periodic orbit lower branches found, two can be continued to large  $Re$: see Fig.~\ref{Fluctuation and Sheet scales}. These green and blue lower branches satisfy $|\vb U|\sim\sqrt{K(\vb U)} \sim Re^0$ and $|\vb u'| \sim \sqrt{K(\vb u')} \sim Re^{-1/2}$, while the upper green branch satisfies $|\vb U| \sim |\vb u'| \sim Re^0$. We focus on the former case, and demonstrate that the role of the fluctuations is to counteract the damping effect  of viscosity on the oscillating sheets. Motivated by the observed scalings, we substitute the expansions 
\begin{equation}
\vb U = \vb U_0 + Re^{-1}\vb U_1 + O(Re^{-2}), \qquad (\vb u',\eta') = Re^{-1/2}(\vb u'_0,\eta'_0) + O(Re^{-1})
\end{equation}
into equations (\ref{V equation}), (\ref{eta fluctuations equations}) and (\ref{v fluctuations equations}).  
The sheet equation (\ref{V equation}) at $O(Re^0)$ simply recovers the inviscid oscillating sheet equation
\begin{align}\label{V equation Re^0 expansion}
    \left[\frac{\partial}{\partial t} + y \cos(\omega t) \frac{\partial} {\partial x}\right] \nabla^2 V_0 = 0,
\end{align}
where $\vb U_0=U_0 \vb{e}_x+V_0 \vb{e}_y$, which is consistent with the lower branch solutions shown in \cref{New SSP Sheets}(a) and (b) since $|\vb u'|\sim Re^{-1/2}$ and $D_m\ll 1$. To understand how the $M$ amplitudes $A_m$ in the inviscid sheet
\begin{equation}
V_0 = \sum^M_{m=1} \frac{A_m}{k_x^2+k_y(t)^2}\exp \left(im(k_x x+k_y(t)y) \right)+c.c.
\end{equation}
are determined, we turn to the fluctuation equations (\ref{eta fluctuations equations}) and (\ref{v fluctuations equations}) at $O(Re^{-1/2})$,
\begin{align}\label{eta fluctuations equations Re^-1/2}
    \left[\frac{\partial}{\partial t} + y\cos(\omega t) \frac{\partial} {\partial x} \right] \eta'_0+ \frac{\partial v'_0}{\partial z} \cos(\omega t) + \frac{\partial}{\partial z}\left(\u'_0 \cdot \nabla U_0 \right) = 0,
\end{align}
\begin{align}\label{v fluctuations equations Re^-1/2}
    \left[\frac{\partial}{\partial t} + y \cos(\omega t) \frac{\partial} {\partial x} \right] \nabla^2 v'_0 
     + \nabla^2(\u'_0 \cdot \nabla V_0) = 0,
\end{align}
which are linear in the fluctuations, and depend only on the leading order sheet flow $\vb U_0$. For a periodic orbit, the fluctuations must be a neutral Floquet mode, discretising the set of admissable $\vb U_0$ (as not all $\vb U_0$ have neutral Floquet modes, see \cref{linear stability of sheets}). Given this neutrality condition is generically codimension-1, this reduces the $M$ complex unknowns of the sheet down to $2M-1$ real variables.  
To now understand how the fluctuation amplitude is set for a given $\vb U_0$, we turn to the next order in the sheet equation (\ref{V equation}) at $O(Re^{-1})$
\begin{align}\label{V equation Re^-1 expansion}
    \underbrace{\left[\frac{\partial}{\partial t} + y \cos(\omega t) \frac{\partial} {\partial x}\right] \nabla^2}_{\mathcal{L}} V_1 - \nabla^4 V_0 
    +   \underbrace{\left\langle \frac{\partial}{\partial y}\left(\nabla \cdot(\u'_0 \cdot \nabla \u'_0)\right) - \nabla^2(\u'_0 \cdot \nabla v'_0) \right\rangle_z}_{\mathcal{N}(\vb u'_0)} = 0,
\end{align}
and derive the Fredholm solvability condition. Under the $L_2$ inner product
\begin{equation}
\langle\phi_1,\phi_2\rangle_{x,y,t}=\frac{\omega k_x}{(2\pi)^3}\int_0^{2\pi/\omega}\int^{\pi}_{-\pi} \int^{\pi/k_x}_{-\pi/k_x} \phi_1^* \phi_2\,dx \, dy \,dt
\end{equation}
the adjoint of $\mathcal{L}$ is the operator 
\begin{equation}
\mathcal{L}^\dagger = -\nabla^2\left[\frac{\partial}{\partial t} + y \cos(\omega t) \frac{\partial} {\partial x}\right]
\end{equation}
which has a kernel populated by the functions $\phi_m=\exp\left(im[k_xx+k_y(t)y]\right)$. With this, taking an inner product of $\phi_m$ and (\ref{V equation Re^-1 expansion}), and noting $\langle\phi_m,\mathcal{L}V_1\rangle_{x,y,t}=\langle\mathcal{L}^\dagger\phi_m,V_1\rangle_{x,y,t}=0$ gives
\begin{align}\label{Fredholm Solvability}
      \left\langle \phi_m, - \nabla^4 V_0 
    +   \mathcal{N}(\vb u'_0) \right\rangle_{x,y,t} = 0
\end{align}
for each $m$ or $M$ complex equations which is precisely sufficient to specify the sheet ($2M-1$ real constants) {\em and} one real fluctuation amplitude. 
Given that the $\phi_m$ span all the spanwise-independent Kelvin modes which make up $V_0$, these equations actually imply
that
\begin{equation}
\int^{2 \pi/\omega}_0 -\nabla^4 V_0 + \mathcal{N}(\vb u'_0) \, dt \, =\,0
\label{sheet_balance}
\end{equation}
which indicates that the sheet energy introduced via the Reynolds stresses exactly offsets the viscous dissipation when averaged over one cycle pointwise in space. The balance (\ref{sheet_balance}) was checked numerically for the lower branch solutions in \cref{Fluctuation and Sheet scales} (a) and (b) with correspondence found down to the 4th significant figure in all components (separately for $m=1$ and $m=2$ components in each orbit). 

The upper green branch in \cref{Fluctuation and Sheet scales} has $|\vb u'| \sim \sqrt{K(\vb u')} \sim Re^{0}$ as $Re \rightarrow \infty$. In this case, the leading order sheet equation has a contribution from the stronger fluctuations and so the sheets are not asymptotically close to the inviscid solution. Consideration of the sheet structure on this branch (\cref{New SSP Sheets}(c)) shows large deviations from the inviscid sheet in the subharmonic component, with $D_2=4.5$, although the harmonic sheet component remains close in shape to the inviscid solution with $D_1=0.1$. In this case, there is no simple asymptotic separation of the component parts.

\subsection{The energising processes}\label{energising processes}

We now discuss how energy is introduced and dissipated for both the sheets and fluctuations for the lower branch solutions where $|\vb u'| \sim Re^{-1/2}$. For the sheets the energy balance is captured succinctly by (\ref{sheet_balance}), which shows that the time-averaged viscous stress is balanced by the Reynolds stresses exerted by the fluctuations.

To understand the energising process for the fluctuations, we consider the fluctuation equations linearised about the sheets (\ref{eta fluctuations equations linearised}) and (\ref{v fluctuations equations linearised}) and name the terms:
\begin{align}\label{eta fluctuations equations mechs named}
    \biggl[\frac{\partial}{\partial t} + \underbrace{y\cos(\omega t) \frac{\partial} {\partial x}}_{\text{background Orr}}-\frac{1}{Re}\nabla^2 \biggr] \eta'+ \underbrace{\frac{\partial v'}{\partial z} \cos(\omega t)}_{\text{background lift up}} + \underbrace{\frac{\partial}{\partial z}\left(u' \partial_x U \right)}_{\text{sheet push forward}} + \underbrace{\frac{\partial}{\partial z}\left(v' \partial_y U \right)}_{\text{sheet lift up}}= 0,
\end{align}
\begin{align}\label{v fluctuations equations mechs named}
    \bigg[\frac{\partial}{\partial t} + \underbrace{y \cos(\omega t) \frac{\partial} {\partial x}}_{\text{background Orr}}-\frac{1}{Re}\nabla^2 \biggr] \nabla^2 v' 
    +   \underbrace{ \nabla^2(u' \partial_x V)}_{\text{sheet push forward}}+   \underbrace{ \nabla^2(v' \partial_y V)}_{\text{sheet lift up}} =0.
\end{align}

%
%
\begin{table}
    \centering
    \begin{tabular}{cc}
Name & Origin \\
 \hline
 Background Orr & $U_B \partial_x\vb{u}'$ \\  
 Background lift up & $v'\partial_y \vb U_B$  \\
 Sheet push forward & $u'\partial_x \vb U$ \\
 Sheet lift up & $v'\partial_y \vb U$ \\
    \end{tabular}
\caption{For each process named in equations (\ref{eta fluctuations equations mechs named}) and (\ref{v fluctuations equations mechs named}), we show which term in the original Navier-Stokes nonlinearity produces it.}
\label{possible mechanisms}
\end{table}

Each named term originates from a different part of the Navier-Stokes nonlinearity $(\vb U_B + \vb U + \vb u') \cdot \nabla (\vb U_B + \vb U + \vb u')$, as summarised in \cref{possible mechanisms}. We introduce `push forward', in which the streamwise fluctuation velocity pushes a base shear forward in $x$, in analogy to lift up \citep{Ellingsen1975} and push over \citep{Lozano21} where the other fluctuation velocity components shift base shears in $y$ and $z$. We also distinguish between background and sheet processes, depending on which base shear is relevant. 

\begin{figure}
\centering
\includegraphics[width=0.8\linewidth]{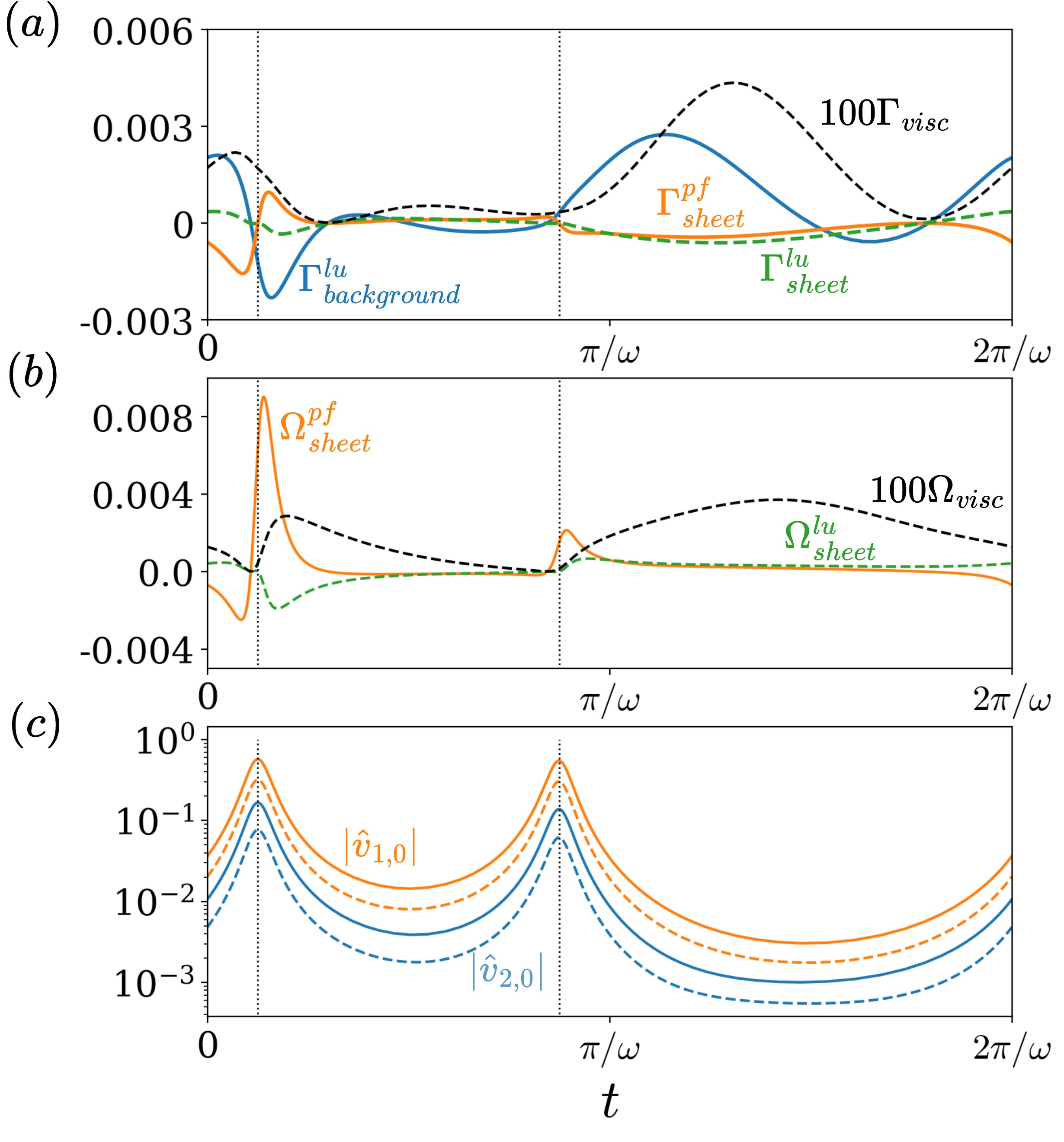} 
\caption{The energy budget for the SSP of \cref{SSP_timeseries} ($Re=1000$, $\omega=0.1$, $k_z=5$, $k_x=0.26$ with $M=N=2$) showing (a) $\Gamma^{lu}_{background}$, $\Gamma^{pf}_{sheet}$, $\Gamma^{lu}_{sheet}$, $\Gamma_{visc}$ and (b) $\Omega^{pf}_{sheet}$, $\Omega^{lu}_{sheet}$, $\Omega_{visc}$. 
(c) The $m=1$ (orange) and $m=2$ (blue) sheet components of the SSP converged when only background lift up and sheet push forward energise the fluctuations (solid) alongside the sheets of the SSP from \cref{SSP_timeseries} (dashed).} 
\label{energising mechs}
\end{figure}

We now show that background lift-up and sheet push forward are the key energising processes sustaining the fluctuations. By multiplying \cref{eta fluctuations equations mechs named} by $\eta'$ and \cref{v fluctuations equations mechs named} by $\nabla ^ 2 v'$  and integrating both over the volume we obtain 
\begin{align}
    \frac{1}{2}\frac{\partial}{\partial t} \langle\eta'^2\rangle_{xyz} &= \Gamma^{lu}_{background} + \Gamma^{pf}_{sheet} + \Gamma^{lu}_{sheet} - \Gamma_{visc}\label{eta energy}\\
    \frac{1}{2}\frac{\partial}{\partial t}  \langle(\nabla^2 v')^2 \rangle_{xyz} 
    &=\Omega^{pf}_{sheet} + \Omega^{lu}_{sheet}- \Omega_{visc}.\label{v energy}
\end{align}
where 
\begin{align}\label{energy terms}
    \Gamma^{lu}_{background} &= - \left\langle\eta'\frac{\partial v'}{\partial z}\cos(\omega t)\right\rangle_{xyz}, \qquad
    \Gamma^{pf}_{sheet} =- \left\langle\eta'\frac{\partial}{\partial z}(u' \partial_x U )\right\rangle_{xyz}\\
    \Gamma^{lu}_{sheet} &=- \left\langle\eta'\frac{\partial}{\partial z}(v' \partial_y U )\right\rangle_{xyz}, \qquad
    \Gamma_{visc} = \frac{1}{Re}\left\langle|\nabla \eta|^2\right\rangle_{xyz}\\
    \Omega^{pf}_{sheet} &=-    \left\langle(\nabla^2 v')\nabla^2(u' \partial_x V) \right\rangle_{xyz}, \qquad
    \Omega^{lu}_{sheet} = -    \left\langle(\nabla^2 v')\nabla^2(v' \partial_y V)\right\rangle_{xyz}\\
    \Omega_{visc} &=   \frac{1}{Re}\left\langle|\nabla (\nabla^2 v')|^2\right\rangle_{xyz}
\end{align}
which correspond to, in order, the $\eta'$ fluctuation energy via background lift up, sheet push forward, sheet lift up, and viscous dissipation, and the $v'$ fluctuation energy via sheet push forward, sheet lift up, and viscous dissipation. The background Orr terms does not contribute to the energy budget, as it vanishes on integrating due to the periodic boundary conditions. For the periodic orbit shown in \cref{SSP_timeseries}, we show how each of these terms evolves with time in \cref{energising mechs}(a) and (b). While viscosity always dissipates energy ($\Gamma_{visc},\Omega_{visc}>0$), it is much smaller in magnitude than the other terms. This is consistent with viscous dissipation not appearing to leading order in the fluctuation equations (\ref{eta fluctuations equations Re^-1/2}) and (\ref{v fluctuations equations Re^-1/2}). All other terms change sign throughout the SSP, showing that they add and remove energy at different points in the cycle.
Background lift up is generally the largest contributor to the right hand side of \cref{eta energy} (see blue line in \cref{energising mechs}(a)), while sheet push forward is the largest contributor for \cref{v energy} (see orange line in \cref{energising mechs}(b)). 

To test if these are really all that is needed, we tried converging a periodic orbit where only these 2 energising terms were kept by treating the reduced fluctuations equations
\begin{align}\label{eta fluctuations equations reduced}
    \biggl[\frac{\partial}{\partial t} + \underbrace{y\cos(\omega t) \frac{\partial} {\partial x}}_{\text{background Orr}} - \frac{1}{Re}\nabla^2\biggr] \eta'+ \underbrace{\frac{\partial v'}{\partial z} \cos(\omega t)}_{\text{background lift up}}=0,
\end{align}
\begin{align}\label{v fluctuations equations reduced}
    \biggl[\frac{\partial}{\partial t} + \underbrace{y \cos(\omega t) \frac{\partial} {\partial x}}_{\text{background Orr}} - \frac{1}{Re}\nabla^2 \biggr] \nabla^2 v' 
    +   \underbrace{ \nabla^2(u' \partial_x V)}_{\text{sheet push forward}} =0
\end{align}
together with \cref{V equation}. The sheet components of the resulting periodic orbit are shown in \cref{energising mechs}(c). When either the background lift up or sheet push forward are also removed no periodic orbit was found. In fact, the only way that the imposed background shear can energise the system is via the background lift up. Ideally, we would also check whether the background Orr terms are required. However, this is not simple as these terms are automatically removed when Kelvin modes are used to solve the system. These Orr terms do not contribute to the energy budget but may still be crucial to the dynamics.

%
%

%
%

\section{Discussion}\label{discussion}

In this paper we have identified a new self-sustaining process in a system driven by a temporally-oscillating, spatially-homogeneous shear. Instantaneously, this SSP is 2D with a time-dependent invariant direction in the cross-shear plane. It consists of oscillating `sheets' that are additionally spanwise-independent (making them instantaneously 1D), and spanwise-dependent fluctuations. The sheets are Euler solutions which would secularly decay in the presence of viscosity. However, in the SSP, a linear instability of the sheets generates fluctuations, which self-interact nonlinearly to re-energise the sheets against viscosity.  The linear instability of the sheet requires the familiar lift up of the imposed oscillating shear and a `push-forward' mechanism due to the streamwise shear introduced by the sheets. 

This new form of SSP differs substantially from the now-well known roll-streak-wave SSP \citep{Hamilton1995, Waleffe1997, Hall10} that is generic to steady shear flows and recently uncovered in a modified form in  Stokes's second problem \citep{SandovalEaves25}. In this, streamwise (2D) rolls generate streamwise-invariant (2D) streaks by lift-up which, when sufficiently strong, become linearly unstable to 3D waves. These close the cycle of energy transfer by nonlinearly-self interacting to sustain the rolls.   We did not identify any of these structures within our model, presumably as the Kelvin mode ansatz used here enforces invariance in the $-k_y(t)\vb{e}_{x}+ k_x\vb{e}_{ y}$ direction, meaning streamwise-independent 2D structures are precluded. To capture these, this enforced invariance in the ansatz has to be broken by including more initial cross-shear wavenumbers than the single one assumed here in our `minimal' representation. It is also worth remarking that in this initial work, we have generally made the simplest choices in the interests of clarity. For example, only spatially-harmonic flows having the same spanwise periodicity as the imposed streaks are considered  and only spatially-harmonic instabilities of the sheets have been used to construct orbits (both temporally harmonic and subharmonic flows are shown in Fig.~\ref{Fluctuation and Sheet scales}).

The periodic orbits found here are certainly relevant to studies which formally impose periodic shear to examine its effects \citep[e.g. on homogeneous turbulence, ][]{Yu2006, Hamlington2009} but the situation is not immediately  clear for bounded flows which are either forced by oscillating boundaries or body forces. Perhaps the most canonical example is a bounded version of Stokes's second problem: oscillatory plane Couette flow where the fluid in the layer $(x,y) \in {\mathbb R} \times [-h,h]$ is forced by the boundaries $y= \pm h$ oscillating with velocity $\pm U_0 \cos(\omega t) \vb{e}_x$. Here the 1D flow response is
\begin{equation}
U(y,t)= \Re \!\left[U_0 
\frac{\sinh (\lambda y)}{\sinh (\lambda h)}e^{i \omega t} \right] \qquad {\rm with}\, \, \,\lambda:= \lambda_r+i \lambda_i=(1+i)\sqrt{ \frac{\omega}{2 \nu} }
\end{equation}
which has an oscillating, approximately spatially-constant shear in the interior mimicking that studied here. Estimating the size of this shear by
\begin{equation}
S^*:=\max_t \frac{\partial U}{\partial y}\biggl|_{y=0}\biggr.=\frac{U_0}{h}\left| \frac{\lambda h}{\sinh (\lambda h)} \right|
\end{equation}
and the extent over which it maintains the same sign by the interval $[-y^*,y^*]$ where $y^*$ is the smallest root of 
\begin{equation}
\Re \left[ \cosh (\lambda y^*) \right]=\cosh (\lambda_r y^*) \cos (\lambda_i y^*) =0
\end{equation}
(the condition for $\partial U/\partial y=0$ at the time corresponding to the largest shear at $y=0$) suggests a length scale 
\begin{equation}
L^* := 2y^*= \frac{\pi}{\lambda_i}.
\end{equation}
These then allow a local Reynolds number based on the shear (as defined in (\ref{Re}) and taking $L^*=2\pi/k_y(0)$\,)  and the shear-normalised oscillation frequency to be estimated as 
\begin{align}
%
%
Re^* &:= \frac{ S^* L^{*2} }{4 \pi^2 \nu}= 
\frac{U_0}{h} \left| \frac{\lambda h}{\sinh (\lambda h)} \right| \frac{1}{4\lambda_i^2 \nu}=
\frac{1}{2St}
\left| \frac{ (1+i)\sqrt{StRe/2} }{\sinh [ (1+i)\sqrt{StRe/2} ]} \right|, \\
%
%
\omega^* & := \frac{\omega}{S^*}= St \left| \frac{\sinh [ (1+i)\sqrt{StRe/2} ]}{ (1+i)\sqrt{StRe/2} } \right|
\end{align}
where the Strouhal number $St$ and `global' Reynolds number $Re$ based on the plate speed are
\begin{equation}
St:= h \omega/U_0, \qquad Re:=hU_0/\nu.
\end{equation}
These make it clear that the local Reynolds number $Re^*$ can be made as large as required by adjusting the Strouhal number $St$ (together with the global Reynolds number $Re$ if the shear is also to be maintained) but the shear-normalised oscillation frequency always  changes to compensate so that
\begin{equation}
\omega^* Re^* = \frac{\omega (L^{*}/2\pi)^2}{\nu}= \tfrac{1}{2}
\end{equation}
i.e. the Reynolds number based on the oscillation frequency is $O(1)$. The finite-amplitude periodic orbits displayed in Fig.~\ref{Fluctuation and Sheet scales} have $\omega^*=0.1$ and $Re^* \gtrsim 500$ and so a larger product of  $50$. It is then not obvious that the periodic orbits found in  the unbounded analysis could manifest in the interior of oscillating plane Couette flow. A full investigation of parameter space may well reveal periodic orbits at lower $\omega^*$ but this has  not been pursued here.

More generally, the processes uncovered here may extend straightforwardly to oscillating shears of a more complicated spatial structure so increasing the effective length scale of the shear in, for example, oscillating plane Couette flow. To investigate this, a similar procedure (embedding in a more complicated situation to secure a bifurcation and then using homotopy to trace back to the original problem) could be used of course but it will be much more expensive to pursue as the Kevin mode simplication is lost.

In terms of future work, the obvious first issue is to establish the relevance of the finite amplitude solutions found here to wall-bounded flows. If they can be found, then all sorts of interesting questions arise as to whether they can co-exist with standard SSP solutions in situations where the forcing has both a steady and oscillatory part, a situation where there are many applications and hence interest.

\FloatBarrier 

\vspace{1cm}
\noindent
{\em Acknowledgements}. T.A.L. is very grateful to the Cambridge Trust for their support.

%
\appendix

\section{Newton-GMRES}\label{GMRES}

For systems of the form (\ref{ODE form}), periodic orbits are converged using the Newton-GMRES algorithm. Using the vector $\phi$ to represent the fluid state, an orbit of period $T$ is equivalent to finding a zero of 
$$F( \phi) \coloneq \psi_T( \phi) -  \phi,$$ 
where recall $\psi_T(\phi)$ corresponds to timestepping $\phi$ into the future $T$ time units. Starting with an initial guess $\phi_0$, we will iteratively take Newton steps $\delta \phi$ that lower the residual of $F$ until it is sufficiently close to zero. By Taylor expanding
$$F( \phi_0+\delta \phi) = F( \phi_0)+dF(\phi_0)\delta\phi + O(|\delta\phi|^2),$$
and choosing a step $\delta\phi$ such that $F$ vanishes after the step with $F( \phi_0+\delta \phi)\approx0$, we obtain
\begin{equation}\label{GMRES step}
dF(\phi_0)\delta\phi = -  F( \phi_0).
\end{equation}
If we had access to the full Jacobian $dF(\phi_0)$, this could simply be inverted to obtain $\delta\phi$ but we don't. Instead, the GMRES algorithm is used which iteratively solves matrix equations $Ax=b$ with matrix $A\in\mathbb{R}^{N\times N}$ and vector $b\in \mathbb{R}^N$ to obtain the vector $x\in \mathbb{R}^N$ \citep{Saad1986}. GMRES requires only access to the action of $A$ on a vector, rather than access to its explicit terms. To solve \cref{GMRES step}, we set $A=dF(\phi_0)$, $b=-  F( \phi_0)$ and $x=\delta\phi$, and we find the action of the Jacobian via central finite difference
$$dF( \phi_0) v = \frac{F(\phi_0+\epsilon v) - F(\phi_0-\epsilon v)}{2\epsilon}$$
for small $\varepsilon$ (here $10^{-8}$). The algorithm was terminated when 
$$\frac{|dF(\phi_0)\delta\phi +  F( \phi_0)|}{|F( \phi_0)|}<10^{-5}$$
The Newton-GMRES algorithm therefore involves taking an initial best guess $\phi_0$, solving \cref{GMRES step} using GMRES, and then updating the best guess $\phi_0 \leftarrow\phi_0+\delta\phi$. This is iterated until $F(\phi)<10^{-10}$, corresponding to a $\phi$ that is on a periodic orbit.

A useful adaptation to this is arclength continuation, which allows the convergence of solution branches as a parameter $\lambda$ varies ($\lambda=\beta$ in \cref{arclength continuation} and $\lambda=\log(Re)$ in \cref{Fluctuation and Sheet scales}). To do this, we follow \cite{Chandler_2013} and add an extra dimension to both the state vector and the residual function so that \begin{equation}
\tilde \phi \coloneq 
\begin{pmatrix} \phi\\
\lambda
\end{pmatrix}\quad \tilde F \coloneq 
\begin{pmatrix} F(\phi)\\
\frac{\partial \tilde\phi_{-1}}{\partial r}\cdot(\tilde\phi-\tilde\phi_{-1}) - \alpha^2\delta r_{-1}
\end{pmatrix}
\end{equation}
where $r$ is an arclength parameter, $\delta r_{-1} = |\tilde\phi_{-1}-\tilde\phi_{-2}|$, $\tilde\phi_{-1},\tilde\phi_{-2}$ are previous states on the branch, $\alpha$ controls how large a step to take and 
$$\frac{\partial \tilde\phi_{-1}}{\partial r} = \frac{\tilde\phi_{-1} - \tilde\phi_{-2}}{\delta r_{-1}}$$
is a first-order estimation of the gradient.
When the final component in $\tilde F$ vanishes, we approximately have $|\partial \tilde \phi / \partial r|^2=\alpha^2$, meaning a converged solution has moved along the branch by an arclength controlled by $\alpha$.
Newton-GMRES can be performed on $\tilde F(\tilde \phi)$ in the same way it was performed on $ F( \phi)$, and roots of $\tilde F$ now correspond to periodic orbits that are further along the solution branch from $\phi_{-1}, \phi_{-2}$. Initial guesses for the algorithm typically used second-order gradient estimates
$$\tilde\phi_0 = \tilde\phi_{-1} + \alpha ^2(D\phi_{-1})\delta r_{-1},$$
with 
$$D\phi_{-1}= \frac{1}{\delta r_{-1}\delta r_{-2}(\delta r_{-1}+\delta r_{-2})}\bigg[\big(2\delta r_{-1}+\delta r_{-2}\big)\delta r_{-2} \tilde \phi_{-1} - \big(\delta r_{-1}+\delta r_{-2}\big)^2\tilde \phi_{-2} +\delta r_{-1}^2 \tilde \phi_{-3} \bigg],$$
although sometimes first-order gradients were used
$$D\phi_{-1} = \frac{\partial \tilde\phi_{-1}}{\partial r}.$$
The parameter $\alpha$ was adjusted to be large enough that the branch was continued quickly, while being small enough that the guesses converged.
\section{Governing Equations}\label{equations}
Here we show how the fluctuation evolution equations are solved using Kelvin modes. The Kelvin mode ansatz (\ref{kelvin mode ansatz}) is substituted into the linearised fluctuation equations (\ref{eta fluctuations equations linearised}) and (\ref{v fluctuations equations linearised}) to convert the PDE into an ODE. This yields the fluctuation equations
\begin{align}\label{eta equation with enforced sheets}
    \dot{\hat{\eta}}_{m,n} + \frac{1}{Re}k_{m,n}^2\hat \eta_{m,n}  
    + \underbrace{ink_z \hat v_{m,n}\cos(\omega t)}_{\text{background lift up}}
    - \sum_{p}np(\underbrace{\hat u_{m-p,n}k_x}_{\text{ sheet push forward}} + \underbrace{\hat v_{m-p,n}k_y}_{\text{sheet lift up}})k_z \hat u_{p,0}  = 0,
\end{align}
\begin{align}\label{v equation with enforced sheets}
    - k_{m,n}^2\dot{\hat{v}}_{m,n} &+ \left[2mk_xk_y \cos(\omega t) - \frac{1}{Re}k_{m,n}^4\right] \hat v_{m,n} \nonumber\\ &-  \sum_{p}ipk_{m,n}^2(\underbrace{\hat{u}_{m-p,n}k_x}_{\text{sheet push forward}} + \underbrace{\hat v_{m-p,n}k_y}_{\text{sheet lift up}}) \hat v_{p,0} =0
\end{align}
for $n\neq0$, where $k_{m,n}^2 \coloneq m^2(k_x^2 + k_y^2) + n^2 k_z^2$.

To consider the linear instability of the inviscid sheets (\cref{linear instab of sheet generate fluc}), we enforce that the sheet components are inviscid with $$\hat v_{p,0}=\frac{A_p}{k_x^2+k_y(t)^2}, \quad \hat u_{p,0} = -\frac{k_y(t)}{k_x}\frac{A_p}{k_x^2+k_y(t)^2},$$ where $\hat u_{p,0}$ is found by \cref{kelvin mode u}, and $A_p$ is specified in \cref{linear instab of sheet generate fluc}. Equations (\ref{eta equation with enforced sheets}) and (\ref{v equation with enforced sheets}) with these enforced sheets then constitute a complex system of ODEs in only the fluctuation variables ($\hat v_{m,n},\hat \eta_{m,n}$ for $n\neq 0$).

\bibliographystyle{jfm}
\bibliography{bibliography}
\end{document}

%% file: imports.tex
\usepackage{graphicx} 
\usepackage{physics}
\usepackage{xcolor}
\usepackage{multirow}
\usepackage{cleveref}
\usepackage{array}
\usepackage{placeins}
\usepackage{subcaption}
\usepackage{amsmath,amsfonts,amssymb,stackengine,graphicx}
\usepackage{mathtools}

\newcommand{\note}[1]{{\color{red} #1}}

\newcommand{\ky}{\left(1-\frac{k_x}{\omega}\sin(\omega t)\right)}

\renewcommand{\u}{\vb{u}}